\documentclass[runningheads]{llncs}

\usepackage{eccv}

\usepackage{eccvabbrv}

\usepackage{graphicx}
\usepackage{booktabs}

\usepackage{multirow}
\usepackage[accsupp]{axessibility}  % Improves PDF readability for those with disabilities.

\usepackage{hyperref}

\usepackage{orcidlink}

\begin{document}

\title{Enhancing Recipe Retrieval with Foundation Models: A Data Augmentation Perspective} 

\titlerunning{Enhancing Recipe Retrieval with Foundation Models}


\author{Fangzhou Song\inst{1}$^{\ast}$\orcidlink{0009-0004-5390-4115} \and
Bin Zhu\inst{2}$^{\ast}$\orcidlink{0000-0002-9213-2611} \and
Yanbin Hao\inst{1}$^{\dagger}$ \orcidlink{0000-0002-0695-1566} \and Shuo Wang\inst{1}\orcidlink{0000-0002-4881-9344} }

\renewcommand{\thefootnote}{\fnsymbol{footnote}}
\footnotetext[1]{Co-first author. \texttt{Email: fangzhousong@mail.ustc.edu.cn, binzhu@smu.edu.sg}}
\renewcommand{\thefootnote}{$\dagger$}
\footnotetext[1]{Corresponding author. \texttt{Email: haoyanbin@hotmail.com.}}

\authorrunning{F. Song et al.}

\institute{University of Science and Technology of China \and 
Singapore Management University 
}

\maketitle

%\vspace{-3pt}

\begin{abstract}
% Abstract goes here.

Learning recipe and food image representation in common embedding space is non-trivial but crucial for cross-modal recipe retrieval. In this paper, we propose a new perspective for this problem by utilizing foundation models for data augmentation. Leveraging on the remarkable capabilities of foundation models (i.e., Llama2 and SAM), we propose to augment recipe and food image by extracting alignable information related to the counterpart. Specifically, Llama2 is employed to generate a textual description from the recipe, aiming to capture the visual cues of a food image, and SAM is used to produce image segments that correspond to key ingredients in the recipe. To make full use of the augmented data, we introduce \textbf{D}ata \textbf{A}ugmented \textbf{R}etrieval framework (\textbf{DAR}) to enhance recipe and image representation learning for cross-modal retrieval. We first inject adapter layers to pre-trained CLIP model to reduce computation cost rather than fully fine-tuning all the parameters. In addition, multi-level circle loss is proposed to align the original and augmented data pairs,
% we introduce circle loss to regulate cross-modal embedding space, 
which assigns different penalties for positive and negative pairs. 
% With the extra augmented data from recipe and image, multi-level circle loss is proposed to construct multi-objective modal alignment. 
On the Recipe1M dataset, our DAR outperforms all existing methods by a large margin. Extensive ablation studies validate the effectiveness of each component of DAR. Code is available at \textcolor{magenta}{https://github.com/Noah888/DAR}.
% Nowadays, healthy diet is increasingly emphasized, and cross-modal recipe retrieval has attracted much attention as an important task in food computing. In this paper, we propose a paradigm network for cross-modal recipe retrieval using foundation models: \textbf{C}onsolidation, \textbf{A}ugmentation and \textbf{R}egulation (\textbf{CAR}), aiming to align two modalities in the common semantic space. Specifically, we further consolidate the foundation model's semantic understanding of the food domain to obtain better image and recipe embeddings. And  for the challenges of high redundancy of recipe  and weak correlation between images and recipes faced by the recipe retrieval task,  we extract the modal auxiliary information to augment the modal consistency by foundation models. Finally, a multi-circle loss function is proposed to regularize the common space of multiple embeddings. With the full combination of  task characterization and foundation model, our \textbf{CAR} network outperforms the state-of-the-art approach on the benchmark recipe1M dataset, and the ablation prove that each component of the \textbf{CAR} is essential.
  %The abstract should concisely summarize the contents of the paper. 
  %While there is no fixed length restriction for the abstract, it is recommended to limit your %abstract to approximately 150 words.
  %Please include keywords as in the example below. 
  %This is required for papers in LNCS proceedings.
  \keywords{Recipe retrieval \and Data augmentation \and Foundation models}
\end{abstract}

\section{Introduction}
\label{sec:intro}

With the rapidly increasing amount of multimodal food data (e.g., recipes, food images and cooking videos) from various sources, the demand for food computing~\cite{min2019survey} to analyze the food data has grown, ranging from food recognition~\cite{chen2020study, min2023large, liu2024canteen}, cross-modal recipe retrieval~\cite{salvador2017learning, salvador2021revamping,zan2020sentence}, food recommendation~\cite{wang2021market2dish} and food logging~\cite{ming2018food, sahoo2019foodai}. In this paper, 
% we study one of the most significant tasks, 
we focus on cross-modal recipe retrieval, which aims to search for recipe given a food image as query and vice versa.

Great progress has been achieved in cross-modal recipe retrieval by advancing the architectures from convolutional neural network and LSTM~\cite{salvador2017learning, wang2019learning, zhu2019r2gan, zhu2020cross} to transformer~\cite{salvador2021revamping} and pre-trained vision and language models~\cite{shukor2022transformer, voutharoja2023malm,papadopoulos2022learning}.
% \textcolor{red}{ We seek to achieve better retrieval performance by learning from paired recipe-image data using a better general network design and more learnable parameters. It is as if we are caught in a trade-off between retrieval performance and learning cost. However, the effect of cross-modal retrieval sometimes depends not only on learning from training samples, but perhaps a more clearly aligned cross-modal image-recipe pair will greatly simplify the difficulty of retrieval. The biggest difference between recipe retrieval and general cross-modal retrieval is that there is misaligned information between recipes and corresponding images, which is potentially harmful to retrieval performance. }
While encouraging, these works pay little attention to a fundamental perspective, i.e., the data misalignment. To be specific, a recipe and its corresponding image could contain misaligned information to each other, which is potentially harmful to retrieval performance.
On the one hand, a recipe is a long document describing the process of cooking a dish, while a food image is the consequence of following the recipe to cook. As a result, the recipe usually contains redundant information that is irrelevant to the visual appearance of the corresponding image, e.g., ``\textit{preheat the oven $350^0$F}'' and ``\textit{Marinate the chicken in the refrigerator for 1 hour}''. On the other hand, the food image usually contains irrelevant information to the recipe as well, such as the background and the food container, etc. 

% Use figure* for multi-column figure
\begin{figure}[t]
    \centering
    \includegraphics[width=0.85\linewidth]{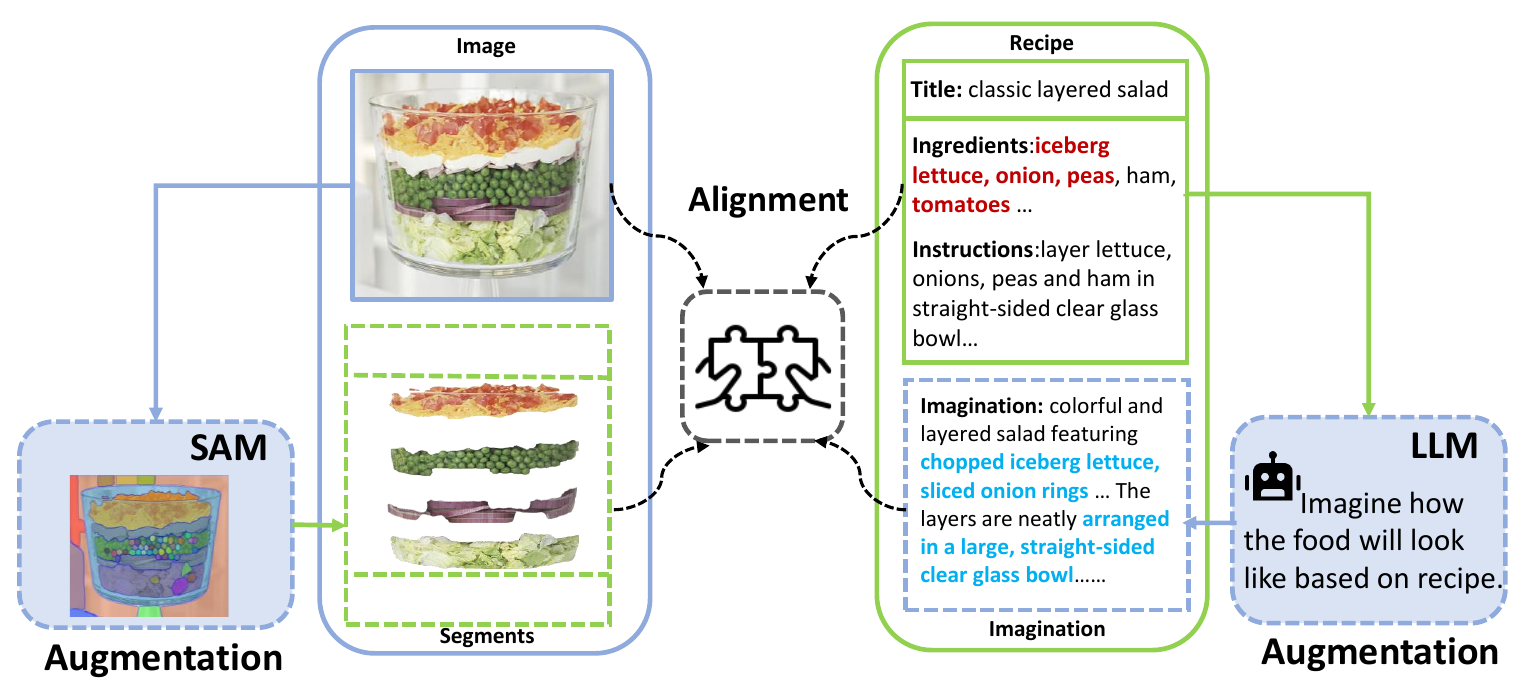}
    
    % \captionsetup{justification=centering}
    \caption{Illustration of our proposed data augmentation paradigm using foundation models. LLM generates textual descriptions from the recipe to capture the dish's visual cues, while SAM produces image segments aligned with recipe ingredients. }
    % Cross-modal recipe retrieval performance is enhanced with the augmented data.}
    \vspace{-0.5cm}
    \label{fig:head}
\end{figure}

This paper proposes a novel data augmentation paradigm by leveraging foundation models to address the limitation. Inspired by the remarkable capabilities of foundation models~\cite{touvron2023llama, ma2023segment, achiam2023gpt, team2023gemini}, 
% \textcolor{red}{This paper provides a new perspective to address the semantic misalignment between recipes and images: data augmentation through foundation model.  With the generic semantic understanding capability demonstrated by the foundation model on multiple downstream tasks, it can compensate for the semantic inconsistencies that exist between the original data and mine information that is more directly related to another modality to be used as auxiliary data, which will be more conducive to cross-modal recipe retrieval.
we introduce language and visual foundation models to augment the recipe and food image data, respectively. Specifically, as shown in Fig.~\ref{fig:head}, to augment the recipe, large language model Llama2~\cite{touvron2023llama} is instructed to act as a helpful assistant to generate a description of the visual appearance of the dish based on recipe as input, which we call ``visual imagination description''. Compared to the original recipe, the visual imagination description concentrates on the content relevant to the visual appearance of a dish image, eliminating the redundant and cumbersome information in the recipe.
% provides a description of the content of the recipe from a visual perspective, which is closer to the food image. 
To augment food image, segment anything (SAM)~\cite{ma2023segment} is utilized to segment the image into multiple segments, which usually correspond to the key ingredients in the recipe. In this way, a fine-grained relationship between food image segments and ingredients in the recipe could be established for cross-modal alignment, while alleviating the impact of irrelevant visual information, such as the background in the image. It is worth noting that the augmented image segments and visual imagination description can be used not only for training but also for evaluation.
% In this way, a fine-grained cross-modal matching relationship between food image segments and recipe's ingredients is established while removing retrieval-irrelevant information such as background from the image.

We propose a Data Augmented Retrieval (DAR) framework that utilizes data augmentation to enhance cross-modal recipe retrieval performance.
% A Data Augmented Retrieval (DAR) framework is proposed to make use of the augmented data to enhance cross-modal recipe retrieval performance.
% \textcolor{red}{Faced with such a wealth of modal information, a framework has been proposed to make better use of the generated auxiliary data to enhance recipe retrieval, called CAR.} 
% first, we introduce adapter layers~\cite{houlsby2019parameter} to consolidate CLIP for cross-modal recipe retrieval. 
We aim to take advantage of the powerful ability of pre-trained CLIP
to encode original image-recipe pairs as well as augmented data with efficient computational cost. As a result,
instead of fully fine-tuning the CLIP model as in~\cite{shukor2022transformer, voutharoja2023malm}, we keep the parameters of CLIP frozen and inject lightweight adapter layers~\cite{houlsby2019parameter} into the CLIP. 
% As a result, We can take advantage of the powerful ability of pre-trained CLIP
% to encode raw image-recipe pairs as well as augmentation data with much less computational cost. 
In addition, 
% in order to better learn the correlation between multiple embeddings, 
we propose a multi-level circle loss to learn cross-modal alignment across original and augmented data during training. The multi-level circle loss is based on the circle loss~\cite{sun2020circle}, which assigns different penalties for positive and negative pairs, thus making the training more flexible compared to the widely used triplet loss. The circle loss is adopted to both original and augmented data pairs.
% not only adopted for original image-recipe pairs, but also employed for augmented image segments and recipe, food image and visual imagination description as well as any two sections of a recipe. 
Consequently, our DAR manages to make full use of the data augmented by the foundation model and achieve state-of-the-art retrieval performance on Recipe1M~\cite{salvador2017learning} dataset. 

In summary, our contributions are as follows:
%\begin{itemize}[left=2em]
% \textcolor{red}{
\begin{itemize}
% \item We propose a new perspective to address the semantic misalignment of images and recipes for the recipe retrieval task by utilizing a foundation model to augment the data. the SAM segments out ingredients objects in the images and the LLM captures visual cues in the recipes. The enhanced data can be used not only in the training phase but also in the evaluation phase.
\item We propose a new data augmentation paradigm for cross-modal recipe retrieval with foundation models. The recipe is augmented as visual imagination description to focus on visual cues by LLM, and the food image is augmented as segments capturing the key ingredients by SAM. The augmentation can be adopted in both training and testing phrases.
\item We introduce Data Augmented Retrieval framework to fully utilize the augmented data using CLIP encoder with lightweight adapter layers and more flexible multi-level circle loss for cross-modal alignment.
\item Our proposed model outperforms existing methods by a large margin on the Recipe1M~\cite{salvador2017learning} dataset. Extensive ablation studies verify the effectiveness of the proposed techniques.
\end{itemize}

\section{Related Work}
\label{sec:related}

\subsection{Cross-Modal Recipe Retrieval} 
% \todo{replace\cite{} with ~\cite{}, apply to the full paper}
% Food computing~\cite{min2019survey} is an area that has recently gained attention around web-rich food data, with a multitude of tasks including food classification~\cite{martinel2018wide} and recognition~\cite{chen2017cross}, recipe recommendation~\cite{wang2021market2dish}  and cross-modal recipe retrieval~\cite{li2021cross,pham2021chef}. This paper focuses on the cross-modal recipe retrieval task. In recent years, many datasets~\cite{bossard2014food,farinella2014benchmark,marin2021recipe1m+,salvador2017learning} for food computing have been proposed, among which Recipe1M~\cite{salvador2017learning} and Recipe1M+~\cite{marin2021recipe1m+} with over 1m cooking recipes and 800k food images are the largest ones. In the face of massive food data, 
Cross-modal recipe retrieval aims at mutual retrieval between recipes and corresponding food images. 
% Cross-modal recipe retrieval is a cross-modal retrieval task focusing on food data.
% Compared to traditional retrieval tasks, the recipe is composed of three structured parts: title, ingredients, and instructions. As the recipe describes the process of how to cook a dish, the corresponding food image is implicitly aligned with the recipe. What’s more, ingredients and instructions texts are generally too long. These characteristics make recipe retrieval more difficult. 
% For task characterization, 
Earlier works~\cite{carvalho2018cross,salvador2017learning,zhu2020cross, xie2021learning,li2021cross, zhu2021learning, pham2021chef} use a pre-trained textual representation (e.g., word2vec~\cite{mikolov2013efficient} for word embedding, skip-thought~\cite{kiros2015skip} for sentence embedding), followed by LSTM to obtain the final recipe embeddings. 
% Among them, \cite{zhu2019r2gan} explored the feasibility of generating image from procedure text for retrieval problem using generative adversarial networks. 
\cite{wang2022paired} introduces StyleGAN2~\cite{Karras_2020_CVPR} to generate images and aligns them in latent space with text for data enhancement. The recipe retrieval performance is further advanced by
% With the Transformer structure demonstrating token relevance capture capability in vision and NLP tasks, some 
transformer-based works~\cite{guerrero2021cross,salvador2021revamping,shukor2023vision,wang2022learning}. 
% \cite{salvador2021revamping} proposes a recipe encoder based on hierarchical Transformers, which encodes the ingredients and instructions text into sentence-level embeddings before obtaining the final embedding, and implements a simple end-to-end embedding learning framework in conjunction with the image encoder. 
% It is followed in many later works~\cite{shukor2023vision,shukor2022transformer,voutharoja2023malm}. 
% For comparison, we also adopt this framework as our baseline. 
\cite{shukor2023vision,shukor2022transformer,voutharoja2023malm} have introduced CLIP-based models and achieved promising performance, but they all fully fine-tuned the image encoder of the pre-trained CLIP to encode food images with extensive computation cost. Different from existing methods, we propose a new data augmentation paradigm to enhance the cross-modal recipe retrieval performance.  
% Accordingly, we propose our paradigm of deeply utilizing multiple foundation models and obtain better retrieval performance.

\subsection{Foundation Model}
% The learning paradigm of AI has been changing in recent years with the development of deep learning~\cite{lecun2015deep}. In contrast to the traditional paradigm of feature engineering based on generic network architecture such as AlexNet~\cite{krizhevsky2012imagenet}, ResNet~\cite{he2016deep} and Transformer~\cite{vaswani2017attention} for different tasks with different data distributions, 
Large language models (LLMs)~\cite{bommasani2021opportunities,achiam2023gpt,team2023gemini, brown2020language, touvron2023llama} have demonstrated strong language abilities and achieved great success by training with large-scale data.
In computer vision, the recently proposed Segment Anything Model (SAM)~\cite{kirillov2023segment}, a visual foundation model can segment the specified content of an image for a given prompt (e.g., box, point, or mask), thus providing semantic visual information for downstream tasks. 
% In the field of recipe retrieval, \cite{wang2022paired} utilizes randomly replaced text and StyleGAN2~\cite{Karras_2020_CVPR} to generate augmented image-text pairs. 
% And for foundation model, it is also widely used for image/text task augmentation. 
These foundation models have been attempted for data augmentation. For example, SAM is used in \cite{zhang2023input} to obtain priority maps for medical image segmentation enhancement. \cite{whitehouse2023llm} explores data augmentation on multilingual commonsense datasets with powerful instruction-tuned LLMs, and diffusion model is used to enhance image diversity in \cite{trabucco2023effective}.
% While the foundation model is highly transferable, the foundation properties suggest that it can be used as the core of a downstream task but needs to be adapted for the task properties. There have been many works~\cite{yu2023cgt,wang2021actionclip,chen2023rsprompter,ma2023segment} utilizing foundation models on downstream tasks and achieving SOTA performance. 
% This paper focuses on leveraging the capabilities of foundation models for data augmentation. 
% On the one hand, we aim to enhance the pre-trained CLIP model by injecting adapter layers to reduce the computation cost. On the other hand, 
In this paper, we employ Llama2~\cite{touvron2023llama} and SAM to generate visual imagination description from recipe to capture visual cues of food images and  image segments corresponding to ingredients in recipe respectively.

% \subsection{Data Augmentation}

\section{Method}
\label{sec:method}
\subsection{Overview} 
Given an image as a query, image-to-recipe retrieval aims to find the most relevant recipe from a recipe corpus, and vice versa. Assume a set of $N$ image-recipe pairs $\{(i_t, r_t)\}_t^N$ are given, where $i_t$ is a food image and $r_t$ is the corresponding recipe. $i_t \in V$ and $r_t \in R$, where $V$, $R$ represent the visual and recipe modal spaces, respectively. As the image and recipe belong to different modalities, they cannot be directly compared. As a result, the key to cross-modal recipe retrieval is to learn a mapping function $\Psi(V, R) \rightarrow (E_V,E_R)$ that maps recipes and images into a common embedding space for similarity measurement, where $E_V$ and $E_R$ are image and recipe embeddings, respectively.
% As a result, the key of cross-modal recipe retrieval is to learn a mapping function $\Psi(V, R) \rightarrow (E_V, E_R)$, which maps the recipe and image in the common embedding space for similarity measurement. 

\begin{figure*}[t]
    \centering
    \includegraphics[width=\textwidth]{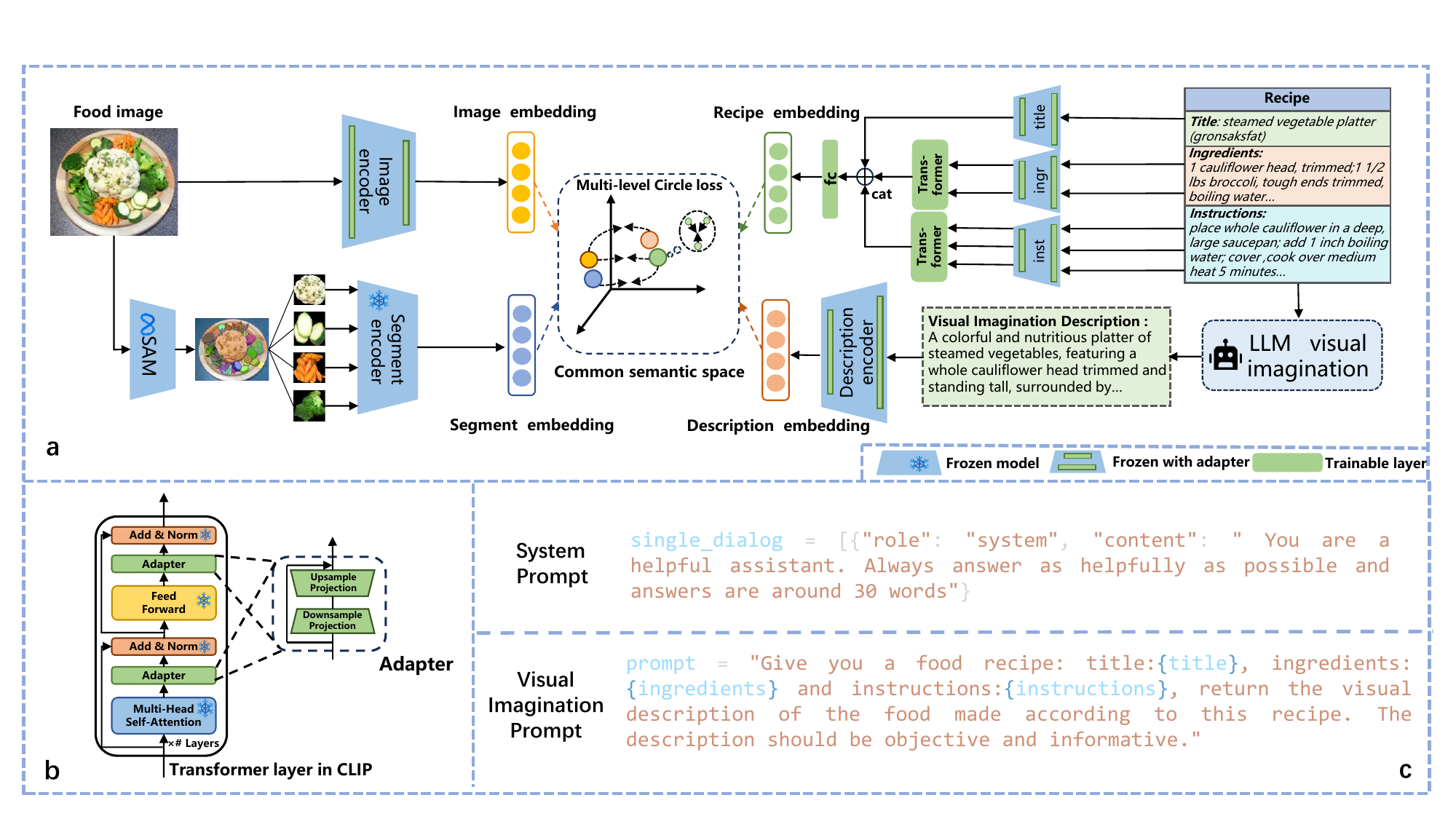}
    %\vspace{-0.8cm}
    % \captionsetup{justification=centering}
    \caption{\textbf{(a)} Overview of the DAR framework architecture. 
    % Adapter layers are integrated to fine-tune the pre-trained CLIP  for cross-modal recipe retrieval. Two auxiliary branches are utilized for data augmentation by the foundation models: image segments via SAM and visual imagination description of the recipes via LLM. The embeddings of the four encoders are aligned in the common space by the multi-level circle loss. 
    \textbf{(b)} Architecture of the adapter in the Transformer layer of CLIP. \textbf{(c)} Prompt of the LLM to generate visual imagination description.}
    \vspace{-0.6cm}
    \label{fig:method}
\end{figure*}

% For cross-modal recipe retrieval, given a set of N image-recipe pairs $(i_t, r_t),t=1,\ldots,N$, the purpose of retrieval is to find the most relevant $r_t$ or $i_t$ to the input $i_t$ or $r_t$, and How to capture this correlation is the key to retrieval. Considering the semantic gap between image and text, it is impractical to directly construct the mapping relation $v \rightarrow r$. In other words, for the domain gap between the visual domain $V(i \in V)$ which $i$  belongs to, and the recipe domain $R( r \in R)$which $r$ belongs to, it is necessary to embed the features of the two domains into the same semantic space. The spatial location relationship between image embedding $E_V$ and recipe embedding $E_R$ then represents the difference in similarity. And  $(V,R) \rightarrow (E_V,E_R)$ mapping function $\Psi(V, R)$ is what the network has to learn.
% \textcolor{red}{
% The goal of this paper is to explore data augmentation paradigm using foundations model to enhance cross-modal recipe retrieval performance.
% leverage foundation models to learn discriminative recipe and image features for cross-modal recipe retrieval.
This paper aims to explore a new data augmentation paradigm using foundation models to enhance cross-modal recipe retrieval performance.
The overview of our proposed model is depicted in Fig.~\ref{fig:method}.  Firstly, we propose to employ foundation models to produce augmented data from recipe and image respectively. Specifically, on the one hand, image segments are extracted from the original food image by the visual foundation model SAM, which corresponds to the key ingredients in the recipe at the semantic level. On the other hand, for the recipe, Llama2 is instructed to imagine the visual description of the food based on the recipe, the result of which we refer to ``visual imagination description'', denoted as $d$. Furthermore, inspired by~\cite{sung2022vl}, we propose to adopt adapters based on pre-trained CLIP model to encode both original and augmented data, then we further introduce a multi-level circle loss function to align original and augmented recipe and image data in the common embedding space.
% }

% \textcolor{red}{
% Next we'll go into more detail about how to augment data with foundation models, and frameworks for utilizing enriched data.}

% Use figure* for multi-column figure

\subsection{Data Augmentation by Foundation Models}
\noindent \textbf{Augment recipe with LLM.}
A recipe contains a list of ingredients and a set of instructions to prepare a particular dish. As a food image is the consequence of the corresponding recipe rather than a caption to describe the image, some information in the recipe is redundant with respect to the visual appearance of the image, for instance, ``\textit{preheat the oven to 400 degrees F}'' and ``\textit{Place yogurt in the strainer and allow it to drain for 15 to 20 minutes}''. The misalignment between recipe and image could hinder cross-modal recipe retrieval. In order to address this issue, unlike previous approaches in designing better encoders for recipe~\cite{wang2022learning,li2021multi}, we propose a new paradigm by employing \textit{Llama2-13b-chat}~\cite{touvron2023llama}, one of the latest Large Language Models (LLM) to extract the visually aligned information from the recipe for textual data augmentation.

% \textcolor{red}{
% Recipes are often truncated in practice due to the length of the text, which undoubtedly results in the loss of key information. In fact, a recipe describes the process of how to utilize ingredients according to the instructions to create a dish named title, some information in a recipe is redundant for the visual appearance of the food image. and thus is independent of cross-modal retrieval. On the other hand, the visual appearance of the food image as a result of a recipe is not explicitly reflected in the recipe, and the visual cues are implicit in the instructions. In order to solve the semantic inconsistency between Recipe and food images, unlike previous approaches in designing the encoder for recipe, we employ \textit{LLama2-13b-chat}~\cite{touvron2023llama}, one of the latest Large Language Models (LLM) to extract the visually aligned information from the recipe for textual data augmentation.
% }

By leveraging the powerful language generation capabilities of the LLM, we aim to produce a description that is visually aligned with a food image, using a recipe as input. As shown in Fig.~\ref{fig:method} (c), we carefully design a ``visual imagination'' prompt to instruct LLM to play the role of a helpful assistant who can imagine what the food will look like based on a recipe. The generated visual imagination description is limited to approximately 30 words to fit the maximum number of input tokens of CLIP. Furthermore, to mitigate the hallucination, we instruct the LLM to generate ``objective'' and ``informative'' responses ``according to this recipe'' in the prompt.

\noindent \textbf{Augment food image with SAM.} In a food image, ingredients are the most informative component, which can be matched to the ingredient section in a recipe. However, irrelevant information with the corresponding recipe is inevitable to be introduced to food images, such as the food plate and background. To mitigate this issue, the visual foundation model SAM~\cite{kirillov2023segment} is adopted to segment the key ingredients from the food image for visual data augmentation.

We use the ``everything''~\footnote{https://segment-anything.com/demo} mode to obtain multiple segments from each food image from SAM. i.e., Zero-Shot object proposal generation. SAM samples a large number of points on the image as the prompt, then filters and de-duplicates the generated prediction masks, and finally generates all the prediction masks for the whole image. Note that not all the segments are useful. For instance, there are a large number of background and decorative noises that are not related to food, and there are also many segments that are too small with less semantic information. As a result, we filter out valuable segments by setting an area threshold and semantically consistent matching of segments with text embeddings of \textit{\{a picture of food\}} using the CLIP model. As shown in Fig.~\ref{fig:method}, the image segments extracted from the food image are the key ingredients to cook the dish. We use the top-$n$ segments as the image augmented data based on the filtering scores. 
% and for those few samples with less than $n$ segments, we believe that a fixed-size randomized crop of the image can also provide some fine-grained ingredients information to make up for the lack of segments.

% \subsection{Framework for Integrating Original and Augmented Data}
\subsection{Data Augmented Retrieval(DAR)}

% \textcolor{red}{
% Now, we add visual imagination description $d$ and and image segments $s$ to the normal raw image data $i$ and recipe data $r$. Faced with such rich data, a framework with low computational cost for multi-embedded input semantic encoders and a more flexible multi-embedded relationship learning strategy, called CAR, is proposed to more fully utilize the augmented data generated by foundation models. and get better retrieval performance.
% }

To fully make use of augmented data, i.e., visual imagination description and image segments, we propose Data Augmented Retrieval (DAR) framework to enhance cross-modal recipe retrieval performance. We aim to make the model efficient without significantly increasing computational cost and introduce a flexible cross-modal alignment learning strategy for the rich set of data.

\noindent \textbf{Fine-tune CLIP with lightweight adapter.} CLIP has demonstrated strong capabilities in various vision-language tasks via contrastive learning on large-scale image-text pairs. There has been some work~\cite{shukor2022transformer, voutharoja2023malm} all fine-tuning CLIP with lots of parameters. To reduce the cost, we propose injecting lightweight adapter layers into the CLIP while keeping all the pre-trained CLIP parameters frozen. As a result, the trainable parameters can be reduced significantly and the adapter layers are trained to enhance the ability of CLIP for cross-modal recipe retrieval. As shown in Fig.~\ref{fig:method} (a), we introduce CLIP image and text encoders with adapter layers to obtain image and recipe embeddings, respectively, i.e., $E_V=\phi_{\text{img\&A}}(i), E_R=\phi_{\text{rec\&A}}(r)$, where $\&A$ stands for adding adapter to the encoders. Fig.~\ref{fig:method} (b) presents the details of one transformer layer in CLIP with adding the adapter layers. Specifically, the adapter consists of a downsampling projection and an upsampling projection, as well as a residual connection.

The CLIP image encoder can be directly applied to encode the food image. In contrast, a recipe is a long document with three parts: title, ingredients and instructions, denoted as $r=(r_{\text{tit}}, r_{\text{ing}}, r_{\text{ins}})$, CLIP text encoder is incapable to model a recipe in one shot. To address this issue, we propose to use three independent CLIP text encoders to encode the three parts separately. As title is usually one sentence, we can directly use title encoder $\phi_{\text{tit\&A}}$ to obtain title features $E_{R_{\text{tit}}}$. Different from title, the ingredients and instructions sections generally consist of multiple sentences, thus we encode each sentence independently with the CLIP text encoder with adapter layers to get the sentence-level features. Moreover, to learn the compact features of ingredients and instructions, we propose to use a two-layer transformer encoder to learn the interactions among the sentences in ingredient and instruction as follows:
\begin{equation}
E_{R_{\text{ing}}} = \textit{Trans}\left(\phi_{\text{ing\&A}}\left([r_{\text{ing}}^1, \ldots, r_{\text{ing}}^{M_{\text{ing}}}]\right)\right), 
\end{equation}
\begin{equation}
E_{R_{\text{ins}}} = \textit{Trans}\left(\phi_{\text{ins\&A}}\left([r_{\text{ins}}^1, \ldots, r_{\text{ins}}^{M_{\text{ins}}}]\right)\right),
\end{equation}
where $M_{\text{ing}}$ and $M_{\text{ins}}$ are the maximum number of sentences that can be acceptable for ingredients and instructions, $[r_*^1,\cdots,r_*^{M_*}]$ form the list of sentences for ingredients or instructions, and $\textit{Trans}$ represents the transformer structure. 

The features of the three parts of the recipe are then concatenated together, which is subsequently fed into a fully-connected (FC) layer to get the final recipe embedding $E_R$, which can be formalized as follows:
\begin{equation}
E_R = Tanh(\textit{FC}(\textit{Concat}(E_{R_{\text{tit}}}, E_{R_{\text{ing}}}, E_{R_{\text{ins}}}))).
\end{equation}

Similar to the operation for the recipe, we employ CLIP text encoder with the adapter to compute the output of the visual imagination description $d$ as follows:
\begin{equation}
    E_D=\phi_{\text{dec\&A}}(d).
\end{equation}

Considering that there are several image segments and to prevent the effect of partially noisy samples from increasing, we use the fully frozen CLIP image encoder $\phi_{\text{seg}}$ to encode each image segment first. The final segment embedding $E_S$ is obtained by averaging the embeddings of the $n$ segments. The formula can be written as:
\begin{equation}
E_S = \frac{1}{n} \sum_{i=1}^{n} \phi_{\text{seg}}^*(s_i),
\end{equation}
where $s_i$ is the image segment and $n$ is the number of segments. $\phi_{\text{seg}}^*$ refers to frozen segment encoder.

\noindent \textbf{Multi-level circle loss with multiple embeddings.} The cross-modal recipe retrieval performance depends on the effectiveness of common embedding space between recipe and image. We propose multi-level alignment using circle loss~\cite{sun2020circle} to regulate cross-modal embedding space based on raw and augmented data, including recipe ($E_R$), food image ($E_V$), image segments ($E_S$) and visual imagination description ($E_D$). 

Existing works typically employ triplet loss and its variants for cross-modal embedding space learning~\cite{zhu2019r2gan,shukor2022transformer,salvador2021revamping,wang2019learning,fu2020mcen}, Given a query, triplet loss aims to push distance between the query and positive samples larger than that of negative pairs with a pre-defined margin, but this means that increasing the cosine similarity score $c_p$ between a query and its positive sample is equivalent to decreasing the cosine similarity score $c_n$ between the query and negative sample, e.g. when $c_p$ is small and $c_n$ approaches 0, it keeps on penalizing $c_n$ with a large gradient, thus the optimization lacks flexibility.

To address this issue, we first introduce circle loss~\cite{sun2020circle} for cross-modal recipe retrieval. Specifically, the key idea is to employ different penalties for $c_p$ and $c_n$ as follows:
%\todo{re-write the definition for circle loss}
%\resizebox{\columnwidth}{!}{$L{_{\text{circle}}(a,b)} = \log \big[ 1 + \sum\limits_{j=1}^{L} \exp(\gamma \alpha_n^j(s{_n^j} - \Delta_n)) \sum\limits_{i=1}^{K} \exp(-\gamma \alpha_p^i(s{_p^i} - \Delta_p))\bigr]$}
\begin{equation}
L_{\text{circle}}(A,B) = \log [ 1 + {\textstyle \sum_{j=1}^{L}} e^{\gamma [ c_{n}^j + m ]_+ \cdot ( c_{n}^j - m ) } {\textstyle \sum_{i=1}^{K}} e^{\gamma [ 1+m- c_{p}^i]_+ \cdot(1-m-c_{p}^i)} ],
\label{eq:circle_loss}
\end{equation}
%L{_{\text{circle}}(a,b)} = \log \big[ 1 + \sum_{j=1}^{L} \exp(\gamma \alpha_n^j(s{_n^j}(a,b) - \Delta_n))\\
%&\times  \sum_{i=1}^{K} \exp(-\gamma \alpha_p^i(s{_p^i}(a,b) - \Delta_p))\bigr],
where $m \in [0,1]$ represents the relaxation factor, and $\gamma$ is the scale factor used to rescale the cosine similarity score. For the set of data pairs $(A, B)$, $K$ and $L$ denote the number of positive and negative pairs for all queries. In addition, to improve the robustness of the model, we use the symmetric bidirectional circle loss, with the following equation:
\begin{equation}  
L(A, B) = L_{\text{circle}}(A, B) + L_{\text{circle}}(B, A).
\end{equation}

We then build the cross-modal embedding space with multiple alignment objectives, including recipe and image $L(E_V,E_R)$, image segment and recipe $L(E_S, E_R)$, and image and visual imagination $L(E_V,E_D)$. Furthermore, as there are three sections in a recipe, i.e., title, ingredients and instructions, we are inspired by ~\cite{salvador2021revamping} to align any two sections of a recipe in a self-supervised manner to encourage semantic consistency within the recipe. We define the recipe loss $L_{\text{rec}}$ using circle loss as follows:
\begin{equation}
L_{\text{rec}} = \frac{1}{6} \sum_{a} \sum_{b} L(E_{R_a}, LN(E_{R_b})) \delta(a, b),
\end{equation}
%\todo{where..., add details here}.
where $a,b\in[tit,ing,ins]$, $\delta (a, b)=1$ if $a\ne b$ otherwise $0$ and $LN(\cdot)$ is a linear projection. $E_{R_a}$ and $E_{R_b}$ are two different content embeddings in a recipe.

Finally, we can obtain the overall multi-level circle loss by combining the above losses as follows:
\begin{equation}
L_{\text{multi-circle}} = L(E_V,E_R) + \alpha L(E_S,E_R)  + \beta L(E_V,E_D) + \sigma L_{\text{rec}},
\end{equation}
where $\alpha$, $\beta$ and $\sigma$ are hyperparameters to balance the losses.

\section{Experiments}
\label{sec:experiments}
\subsection{Setups}

\textbf{Dataset.} In line with previous works, Recipe1M~\cite{salvador2017learning} dataset is used to train and evaluate our method. The recipes with corresponding images are split into 238,408, 51,119 and 51,303 for training, validation and testing respectively. In addition to utilizing these pairs of data, following~\cite{salvador2021revamping}, we also utilize unpaired 482,231 training recipes from the rest of the dataset for the recipe loss $L_{\text{rec}}$.

\noindent \textbf{Evaluation.} Following previous works~\cite{xie2021learning,salvador2017learning, salvador2021revamping}, median rank (medR) and recall rate at top K (R@K) are employed to evaluate the performance of our model. MedR represents the median position of the true positives in the distance ranking in the database, and R@K measures the ranking of the percentage of the top K (with $K \in \{1, 5, 10\}$) containing true positive results. During testing, we randomly sample 1,000 image-recipe pairs and 10,000 image-recipe pairs in the test set as the test subset of two scales ($1k$ set up and $10k$ set up). In the test subset, the embedding of one modality is treated as a query to compute the cosine distance with the candidate embeddings of the other modality in the database, and finally the retrieval results are obtained based on the distance ranking. The final reported results are averaged over the 10 sampled subsets.

For evaluation protocols, the augmented data can be not only used for training but also pre-computed for evaluation. Existing works~\cite{voutharoja2023malm,shukor2023vision} generally compute the cosine similarity between recipe and image as a measure to evaluate the retrieval performance. In contrast, we introduce two extra evaluation protocols based on the augmented visual imagination description and image segments as follows:

% \textcolor{red}{
% Since we introduce auxiliary information visual imagination description $d$ and image segments $s$ for modalities, which can be added not only in training phase, but also in test phase to enhance the performance of the model. For most of the methods~\cite{voutharoja2023malm,shukor2023vision}, take image-to-recipe retrieval as an example, the evaluation protocol is generally to first take the image embedding as a query, compute the cosine distance $dist$ with the recipe embeddings in the database, and then rank the distance to get the retrieval results. We add data augmentation into the distance calculation. Specifically, we set up three evaluation protocols, i.e., CAR, CAR+, and CAR++, as described below:
% }
% \textcolor{red}{
\begin{itemize}
% \vspace{-0.3cm}
  \item \textbf{DAR.} Similar to previous works, DAR only utilizes the distance between image and recipe for evaluation, i.e., $dist=dist_{i-r}$.
  \item \textbf{DAR+} adds visual imagination description into the evaluation, which calculates the distance between image and visual imagination description $dist_{i-d}$, and multiplies by $dist_{i-r}$ to get the distance, i.e., $dist = dist_{i-d} \cdot dist_{i-r}$.
  \item \textbf{DAR++} adds both visual imagination and image segments for evaluation, which computes the distance $dist_{s-r}$ between segments embeddings and recipe embeddings, then multiplies by the distance metric of DAR+, i.e., $dist = dist_{s-r} \cdot dist_{i-d} \cdot dist_{i-r}$.
\end{itemize}
% \vspace{-0.3cm}

\noindent  \textbf{Implementation Details.}
% For the image segments part, we keep 4 image segments after CLIP filtering, and the samples with less than 4 segments are filled with randomly cropped quarter regions of the original image. 
The downsampled dimension of the bottleneck adapter architecture is set to 64. In addition, following~\cite{salvador2021revamping,shukor2022transformer}, the CLIP encoder used in our model is the pre-trained CLIP ViT-B/16 model and the output embedding dimension $d = 512$. For the recipe encoder, ingredients and instructions text can accept a maximum number of sentences $M_{\text{ing}}, M_{\text{ins}}= 15$, and the maximum length of each sentence is 20 tokens. The trainable transformer is 2 layers with 4 attention heads. The dimensionality of the Recipe embedding obtained after the FC layer is also 512.

In the training phase, we set the relaxation factor $m=0.25$ and scale factor $\gamma=32$ for circle loss. The weight factors for multi-level circle loss are set to be $\alpha = 1$, $\beta = 1$, $\sigma = 1$. Following the baseline settings, we train the model with batch size =128 and optimize the parameters using Adam optimizer. The initial learning rate $lr$ is set to $10^{-4}$ and every 30 epochs step decays to 0.1 of the previous $lr$. 
% For the unpaired recipe data in the dataset, considering the small amount of text data, the batch size is set to 256, which is alternately fed into the model with the paired data. 
The model is trained for a total of 100 epochs, and the model parameter with the highest R@1 in validation is selected for testing.

%\usepackage{booktabs}
%\usepackage{multirow}

%\renewcommand{\arraystretch}{1.5}

\begin{table*}[htb]
\centering
\caption{Cross-modal recipe retrieval performance comparison with existing methods. The results are reported in terms of medR ($\downarrow$) and R@K ($\uparrow$). }
% We report three groups of results: ``DAR'' refers to testing with recipe and image same as exiting methods, ``DAR+'' indicates we add augmented visual imagination description, along with recipe and image for testing, and `'DAR++'' represents the reported results are based on adding both augmented image segments and visual imagination description, along with recipe and image for testing.}
\resizebox{\textwidth}{!}{

\begin{tabular}{@{}lcccccccccccccccc@{}}
\toprule
\multicolumn{1}{c}{\multirow{3}{*}{\textbf{Methods}}} & \multicolumn{8}{c}{\textbf{1k}}                                                                                            & \multicolumn{8}{c}{\textbf{10k}}                                                                                           \\ \cmidrule(l){2-17} 
\multicolumn{1}{c}{}                                  & \multicolumn{4}{c}{\textbf{image-to-recipe}}                & \multicolumn{4}{c}{\textbf{recipe-to-image}}                & \multicolumn{4}{c}{\textbf{image-to-recipe}}                & \multicolumn{4}{c}{\textbf{recipe-to-image}}                \\ \cmidrule(l){2-17} 
\multicolumn{1}{c}{}                                  & medR           & R@1           & R@5           & \multicolumn{1}{c}{R@10}          & medR         & R@1           & R@5           & \multicolumn{1}{c}{R@10}          & medR         & R@1           & R@5           & \multicolumn{1}{c}{R@10}          & medR         & R@1           & R@5           & R@10          \\ \midrule
Adamine~\cite{carvalho2018cross}                                                & 2.0          & 40.2          & 68.1          & 78.7          & 2.0          & 39.8          & 69.0          & 77.4          & 13.2         & 14.8          & 34.6          & 46.1          & 14.2         & 14.9          & 35.3          & 45.2          \\
R2GAN~\cite{zhu2019r2gan}                                                  & 2.0          & 39.1          & 71.0          & 81.7          & 2.0          & 40.6          & 72.6          & 83.3          & 13.9         & 13.5          & 33.5          & 44.9          & 12.6         & 14.2          & 35.0          & 46.8          \\
MCEN~\cite{fu2020mcen}                                                   & 2.0          & 48.2          & 75.8          & 83.6          & 1.9          & 48.4          & 76.1          & 83.7          & 7.2          & 20.3          & 43.3          & 54.4          & 6.6          & 21.4          & 44.3          & 55.2          \\
ACME~\cite{wang2019learning}                              & 1.0          & 51.8          & 80.2          & 87.5                      & 1.0          & 52.8          & 80.2          & 97.6                      & 6.7          & 22.9          & 46.8          & 57.9                      & 6.0          & 24.4          & 47.9          & 59.0          \\
SCAN~\cite{wang2022learning}                                                   & 1.0          & 54.0          & 81.7          & 88.8          & 1.0         & 54.9          & 81.9          & 89.0          & 5.9          & 23.7          & 49.3          & 60.6          & 5.1          & 25.3          & 50.6          & 61.6          \\
X-MRS~\cite{guerrero2021cross}                                                  & 1.0          & 64.0          & 88.3          & 92.6          & 1.0          & 63.9          & 87.6          & 92.6          & 3.0          & 32.9          & 60.6          & 71.2          & 3.0          & 33.0          & 60.4          & 70.7          \\
H-T~\cite{salvador2021revamping}                                                    & 1.0          & 60.0          & 87.6          & 92.9          & 1.0          & 60.3          & 87.6          & 93.2          & 4.0          & 27.9          & 56.4          & 68.1          & 4.0          & 28.3          & 56.5          & 68.1          \\
H-T (ViT)~\cite{salvador2021revamping}                                    & 1.0          & 64.2          & 89.1          & 93.4          & 1.0          & 64.5          & 89.3          & 93.8          & 3.0          & 33.5          & 62.1          & 72.8          & 3.0          & 33.7          & 62.2          & 72.7          \\
T-Food (ViT)~\cite{shukor2022transformer}                                           & 1.0          & 68.2          & 87.9          & 91.3          & 1.0          & 68.3          & 87.8          & 91.5          & 2.0          & 40.0          & 67.0          & 75.9          & 2.0          & 41.0          & 67.3          & 75.9          \\
T-Food (CLIP-ViT)~\cite{shukor2022transformer}                                      & 1.0          & 72.3          & 90.7          & 93.4          & 1.0          & 72.6          & 90.6          & 93.4          & 2.0          & 43.4          & 70.7          & 79.7          & 2.0          & 44.6          & 71.2          & 79.7          \\
CREAMY(ViT)~\cite{zou2024creamy}                                                          & 1.0          & 73.3          & 92.5          & 95.6          & 1.0          & 73.2          & 92.5          & 95.8          & 2.0          & 44.6          & 71.6          & 80.4          & 2.0          & 45.0          & 71.4          & 80.0          \\
VLPCook~\cite{shukor2023vision}                                                & 1.0          & 73.6          & 90.5          & 93.3          & 1.0          & 74.7          & 90.7          & 93.2          & 2.0          & 45.3          & 72.4          & 80.8          & 2.0          & 46.4          & 73.1          & 80.9          \\
% VLPCook (R1M+)\cite{shukor2023vision}                                           & 1.0          & 74.9          & 91.4          & 93.7          & 1.0          & 75.6          & 91.2          & 93.6          & 2.0          & 46.7          & 73.3          & 83.3          & 2.0          & 47.8          & 74.1          & 81.8          \\  

FARM~\cite{wahed2024fine}                                                                 & 1.0          & 73.7          & 90.7          & 93.4          & 1.0          & 73.6          & 90.8          & 93.5          & 2.0          & 44.9          & 71.8          & 80.0          & 2.0          & 44.3          & 71.5          & 80.0\\
MALM~\cite{voutharoja2023malm}                                                   & 1.0          & 74.0          & 91.3          & 94.3          & 1.0          & 73.0          & 91.0          & 93.9          & 2.0          & 45.9          & 72.3          & 80.5          & 2.0          & 44.2          & 71.7          & 80.1       \\
CIP(no-Rerank)~\cite{huang2023improving}                                                                  & 1.0          & 73.1         & 94.1          & 97.1          & 1.0          & 72.5          & 93.7          & 97.0          & -          & -          & -          & -           & -           & -           & -            & -           \\
CIP~\cite{huang2023improving}                                                                  & 1.0          & 77.1          & 94.2          & 97.2          & 1.0          & 77.3          & 94.4          & 97.0          & 2.0          & 44.9          & 72.8          & 82.0          & 2.0          & 45.2          & 73.0          & 81.8          \\ \midrule
\textbf{DAR}                                            & \textbf{1.0} & 74.9          & 94.7          & 97.5          & \textbf{1.0} & 75.7          & \textbf{95.4} & \textbf{97.9} & \textbf{2.0} & 44.2          & 73.2          & 82.4          & \textbf{2.0} & 44.8          & 73.9          & 83.1          \\
\textbf{DAR}+                                             & \textbf{1.0} & 76.9          & 94.9          & 97.4          & \textbf{1.0} & \textbf{77.7} & \textbf{95.4} & \textbf{97.9} & \textbf{2.0} & 47.4          & 75.3          & 83.8          & \textbf{2.0} & \textbf{48.3} & \textbf{75.9} & \textbf{84.4} \\
\textbf{DAR}++                                            & \textbf{1.0} & \textbf{77.3} & \textbf{95.3} & \textbf{97.7} & \textbf{1.0} & 77.1          & \textbf{95.4} & \textbf{97.9} & \textbf{2.0} & \textbf{47.8} & \textbf{75.9} & \textbf{84.3} & \textbf{2.0} & 47.4          & 75.5          & 84.1          \\ \bottomrule
\end{tabular}

}

\label{tab:performance_comparison}
\end{table*}

%\vspace{-0.55cm}
\subsection{Performance Comparison}
% \vspace{-0.55cm}
As shown in Table~\ref{tab:performance_comparison}, we compare the cross-modal recipe retrieval performance of our DAR with state-of-the-art methods. Our DAR outperforms all existing methods for the majority of the metrics in both 1k and 10k setups, including CLIP-based methods~\cite{shukor2022transformer, voutharoja2023malm, shukor2023vision, wahed2024fine, huang2023improving}. All of these methods simply fine-tune the entire image encoder of the CLIP to improve the performance. In contrast, by inserting lightweight adapter layers, the number of trainable parameters in our image encoder is only 8\% of theirs. 
Compared to MALM~\cite{voutharoja2023malm}, the image-to-recipe retrieval performance in 1k testing is boosted by 0.9\%, 3.4\% and 3.2\% in terms of R@1, R@5 and R@10 respectively. 
The results show that our DAR learns more discriminative recipe and image embeddings, enhancing cross-modal recipe retrieval through data augmentation.

%T-Food~\cite{shukor2022transformer}, MALM~\cite{voutharoja2023malm} and VLPCook~\cite{shukor2023vision}. All of these methods simply fine-tune the entire image encoder of the CLIP to improve the performance. In contrast, by inserting lightweight adapter layers, the number of trainable parameters in our image encoder is only 8\% of theirs. Compared to VLPCook~\cite{shukor2023vision}, the image-to-recipe retrieval performance in 1k testing is boosted by 1.3\%, 3.8\% and 4.2\% in terms of R@1, R@5 and R@10 respectively. The results show that our DAR learns more discriminative recipe and image embeddings, enhancing cross-modal recipe retrieval through data augmentation.

% The results show that our DAR manages to learn more discriminative recipe and image embeddings for cross-modal recipe retrieval with data augmentation. 

In addition, we also report the results with data augmentation in test time. 
% T
% he second group is to add augmented visual imagination description, along with original recipe and image, referred to ``DAR+''. 
By adding visual imagination description, the retrieval performance of DAR+ can be further improved with noticeable margins in terms of all the metrics than DAR. Furthermore, by adding both augmented image segments and visual imagination description during testing, i.e., ``DAR++'', 
image-to-recipe performance is slightly boosted than DAR+, but the recipe-to-image retrieval performance is inferior to DAR+. We believe the reason is the limitation of image segments, which is analyzed in Sec.\ref{subsection:Qualitative Analysis}. Regarding CLP~\cite{huang2023improving}, the re-ranking step during inference is essential for achieving higher R@1 than our DAR and DAR+, yet it remains inferior to our DAR++. Moreover, our DAR, DAR+ and DAR++ consistently outperform CLP. 

% Comparing CLP~\cite{huang2023improving}, due to its re-ranking strategy in the test phase, we show its results as CLP(no-Rerank) and CLP for a fair comparison. It can be seen from Table~\ref{tab:performance_comparison} that both our original DAR and the DAR++ using data augmentation at test time are respectively superior to it overall.

% \input{tables/ablation_llmprompt}

\subsection{Ablation Study}
%\vspace{-0.3cm}

% \usepackage{multirow}
\begin{table}[]
\centering
\caption{Ablation study for adapters in CLIP encoders for cross-modal recipe retrieval. \&A refers to adding adapter layers to CLIP encoder. The operations are all added to the Zero-shot Retrieval model based on CLIP. All the results are reported in 10K testing with original recipe-image pairs for evaluation.}
\label{tab:ablation_adapter}
\resizebox{\textwidth}{!}{
\begin{tabular}{@{}clcccccccc@{}}
\toprule
\multirow{2}{*}{\textbf{Model}} & \multicolumn{1}{c}{\multirow{2}{*}{\textbf{Operation}}}  & \multicolumn{4}{c}{\textbf{image-to-recipe}}                & \multicolumn{4}{c}{\textbf{recipe-to-image}}                \\ \cmidrule(l){3-10} 
                           & \multicolumn{1}{c}{}                  & \textbf{medR} & \textbf{R@1} & \textbf{R@5} & \textbf{R@10} & \textbf{medR} & \textbf{R@1} & \textbf{R@5} & \textbf{R@10} \\ \midrule
\textbf{Zero-shot Retrieval}        & \multicolumn{1}{c}{}                  & 17.0          & 14.9         & 32.7         & 42.4          & 20.2          & 12.6         & 29.7         & 39.5          \\ \midrule
                           & + \&A to $\phi_{\text{img}}$                     & 5.9           & 24.8         & 49.7         & 61.2          & 4.9           & 27.5         & 53.1         & 64.6          \\
                           & + \&A to $\phi_{\text{rec}}$                     & 4.0           & 28.2         & 55.4         & 66.7          & 4.0           & 28.2         & 55.4         & 66.7          \\
                           & + \&A in $\phi_{\text{img}}$, $\phi_{\text{rec}}$              & 2.0           & 39.2         & 68.7         & 79.0          & 2.0           & 40.7         & 69.7         & 79.8          \\ \midrule
\textbf{Baseline}          & + \&A in $\phi_{\text{img}}$, $\phi_{\text{rec}}$, +  
 $L_{\text{rec}}$      & \textbf{2.0}           & \textbf{42.3}         & \textbf{71.8}         & \textbf{81.5}          & \textbf{2.0}           & \textbf{43.4}         & \textbf{72.8}         & \textbf{82.3}          \\ \bottomrule
\end{tabular}
}

\end{table}

% \usepackage{multirow}
\begin{table}[]
\centering
\caption{Ablation study for augmented data. The operations are all added to the Baseline model. w/ (with) $\phi^*$ and $\phi_{\text{\&A}}$ represent the introduction of augmentation data with a frozen CLIP encoder or a CLIP encoder added to the adapter layers respectively. All the results are reported in 10K testing with original
recipe-image pairs for evaluation.}

\resizebox{\textwidth}{!}{
\begin{tabular}{@{}clcccccccc@{}}
\toprule
\multirow{2}{*}{\textbf{Model}} & \multicolumn{1}{c}{\multirow{2}{*}{\textbf{Operation}}}                                              & \multicolumn{4}{c}{\textbf{image-to-recipe}}                  & \multicolumn{4}{c}{\textbf{recipe-to-image}}                  \\ \cmidrule(l){3-10} 
                                & \multicolumn{1}{c}{}                                                                                 & \textbf{medR} & \textbf{R@1}  & \textbf{R@5}  & \textbf{R@10} & \textbf{medR} & \textbf{R@1}  & \textbf{R@5}  & \textbf{R@10} \\ \midrule
\textbf{Baseline}               & \multicolumn{1}{c}{}                                                                                 & 2.0           & 42.3          & 71.8          & 81.5          & 2.0           & 43.4          & 72.8          & 82.3          \\ \midrule
                                
\multicolumn{1}{l}{}            & + description w/ $\phi_{\text{dec}}^*$                                         & 2.0           & 42.9          & 72.2          & 81.7          & 2.0           & 43.4          & 72.7          & 82.3          \\
                                & + description w/ $\phi_{\text{dec\&A}}$                                                        & 2.0           & 43.7          & 72.6          & 82.0          & 2.0           & 44.7          & 73.9          & 83.1         \\ \midrule
                                & + segments w/  $\phi_{\text{seg}}^*$                                          & 2.0           & 43.0          & 72.3          & 81.9          & 2.0           & 43.6          & 73.0          & 82.6          \\
\multicolumn{1}{l}{}            & + segments w/ $\phi_{\text{seg\&A}}$                                                           & 2.0           & 42.1          & 71.2          & 81.0          & 2.0           & 43.0          & 72.3          & 81.9          \\
                                \midrule
                   & + segment w/  $\phi_{\text{seg}}^*$, description w/   $\phi_{\text{dec\&A}}$ ( \textbf{DAR}) & \textbf{2.0}  & \textbf{44.2} & \textbf{73.2} & \textbf{82.4} & \textbf{2.0}  & \textbf{44.8} & \textbf{73.9} & \textbf{83.1} \\ \bottomrule
\end{tabular}
}

\label{tab:ablation_augmentation}
\end{table}
% \usepackage{booktabs}
% \usepackage{multirow}
\begin{table}[]
\centering
\caption{ Performance comparison with different number of segments from SAM. 
All the results are reported in 10K testing with the DAR++ evaluation.}
\resizebox{0.8\linewidth}{!}{
\begin{tabular}{@{}ccccccccc@{}}
\toprule
\multirow{2}{*}{\textbf{Number of segments}} & \multicolumn{4}{c}{\textbf{image-to-recipe}}                & \multicolumn{4}{c}{\textbf{recipe-to-image}}                \\ \cmidrule(l){2-9} 
                                          & \textbf{medR} & \textbf{R@1} & \textbf{R@5} & \textbf{R@10} & \textbf{medR} & \textbf{R@1} & \textbf{R@5} & \textbf{R@10} \\ \midrule
n = 1                           & 2.0           & 47.7         & 75.5         & 84.1         & 2.0           & 46.2        & 74.6         & 83.7          \\
n = 2                            & 2.0           & 47.7        & 75.7        & \textbf{84.3}          & 2.0           & 46.7         & 75.0         & 84.0          \\
n = 4                            & \textbf{2.0}           & \textbf{47.8}         & \textbf{75.9}         & \textbf{84.3}          & \textbf{2.0}           & \textbf{47.4}         & \textbf{75.5}        & \textbf{84.1}          \\ 
n = 6                            & 2.0           & 47.7        & 75.5        & 84.1          & 2.0           & 46.9       & 75.2       & \textbf{84.1}        \\
\bottomrule
\end{tabular}
}

\label{tab:ablation_segment}
\end{table}
% \usepackage{booktabs}
% \usepackage{multirow}
\begin{table}[]
\centering
\caption{ Performance comparison between circle loss and triplet loss based on Baseline model setup.}
\resizebox{0.8\linewidth}{!}{
\begin{tabular}{@{}ccccccccc@{}}
\toprule
\multirow{2}{*}{\textbf{Loss function}} & \multicolumn{4}{c}{\textbf{image-to-recipe}}                & \multicolumn{4}{c}{\textbf{recipe-to-image}}                \\ \cmidrule(l){2-9} 
                                        & \textbf{medR} & \textbf{R@1} & \textbf{R@5} & \textbf{R@10} & \textbf{medR} & \textbf{R@1} & \textbf{R@5} & \textbf{R@10} \\ \midrule
\textbf{Triplet loss}                            & 2.4           & 36.9         & 66.9         & 77.7          & 2.0             & 37.3         & 67           & 77.8          \\
\textbf{Circle loss}                             & \textbf{2.0}             & \textbf{42.3}         & \textbf{71.8}         & \textbf{81.5}          & \textbf{2.0}             & \textbf{43.4}         & \textbf{72.8}         & \textbf{82.3}          \\ \bottomrule
\end{tabular}
}

\label{tab:ablation_loss}
\end{table}

% In this section, 
We conduct the ablation study to validate the key components of our proposed method DAR, including the adapters in CLIP encoders, augmented data from foundation models and training losses. Unless otherwise specified, all the results are reported in 10K testing for evaluation with original image-recipe pairs.

% \textcolor{red}{
\noindent \textbf{Effect of adapter layers.} Our DAR is built upon CLIP encoders with adapter layers for cross-modal recipe retrieval. As shown in Table~\ref{tab:ablation_adapter}, we firstly show the zero-shot cross-modal recipe retrieval performance using pre-trained CLIP model, where the recipe embedding is averaged over the title, ingredients and instructions embeddings. By adding adapter layers to either CLIP image encoder $\phi_{\text{img}}$ or recipe encoder $\phi_{\text{rec}}$, the retrieval performance is boosted significantly than frozen CLIP. The results show that the introduced adapter layers to CLIP model is effective for cross-modal recipe retrieval. Further gains can be obtained by employing adapter layers to both image and recipe encoders with recipe loss with both paired and unpaired recipes, which is regarded as our baseline model.
% }

\noindent \textbf{Effect of data augmentation by foundation models. } We next examine the effectiveness of augmented data from recipe and image in Table~\ref{tab:ablation_augmentation}. First, retrieval performance improvement can be observed by adding visual imagination description with frozen encoder (i.e., + description w/ $\phi_{\text{dec}}^*$) compared to baseline. The performance is further boosted with adapter layers in the encoder (i.e., + description w/ $\phi_{\text{dec\&A}}$). The results show the effectiveness of visual imagination description for cross-modal recipe retrieval. Second, we further add image segments with frozen encoder (i.e., + segments w/ $\phi_{\text{seg}}^*$) with performance gains. Nevertheless, injecting the adapter layers for segment encoder (i.e., + segments w/ $\phi_{\text{seg\&A}}$) is inferior to frozen one. We believe the reason is the image segments obtained from SAM still contain noise though we have filtered the outputs, which could do harm to 
the segment embeddings with adapter layers. The best performance is achieved by combining both visual imagination description with adapter layers in encoder and image segments with frozen encoder, i.e., our DAR.
% FiDAR.rst, by introducing a frozen segment encoder $\psi_{\text{seg}}$ or an adapter-based description encoder $\psi_{\text{dec\&A}}$ to add the image segment data and visual imagination description, respectively, an improvement in retrieval performance can be observed. The results show that the augmented data incorporated in the respective modalities are effective for cross-modal recipe retrieval. And the results of adding two augmented data at the same time indicate that collaboratively enhancing modal alignment will boost their respective retrieval performance. 

In addition, we conduct experiments by instructing LLM to generate a summary of recipe for data augmentation as well. The results show that visual imagination description manages to outperform summary by 0.7\% and 0.9\% in image-to-recipe and recipe-to-image retrieval respectively. Finally, as listed in Table~\ref{tab:ablation_segment}, we examine the performance with different numbers of segments from SAM. When segments are too few (n=1 or 2), key ingredients may be excluded. Conversely, a large number of segments (n=6) not only raises computational costs but also increases the likelihood of introducing noise. Based on the results, our DAR sets segment numbers to 4.

% \textcolor{red}{Finally, we verified the settings regarding the number of segments retained after SAM segmentation and filtering in Table~\ref{tab:ablation_segment}. We obtained the results of our experiments in the DAR++ protocol of adding image segments into the evaluation, which shows that our choice is superior.}

% we verify the superiority of visual imagination prompt in the Table~\ref{tab:ablation_llmprompt}, in which it outperforms the summary prompt that lets LLM directly generate a summary of the contents of the recipe. And the results further justify the utilization of LLM to enhance the recipe retrieval.

\noindent \textbf{Effect of circle loss.} As shown in Table~\ref{tab:ablation_loss}, we compare the retrieval performance between triplet loss and circle loss under the same setting as our baseline model. Circle loss significantly outperforms the triplet loss across all the metrics. 

% Use figure* for multi-column figure
\begin{figure}[h]
    \centering
    \includegraphics[width=0.86\linewidth]{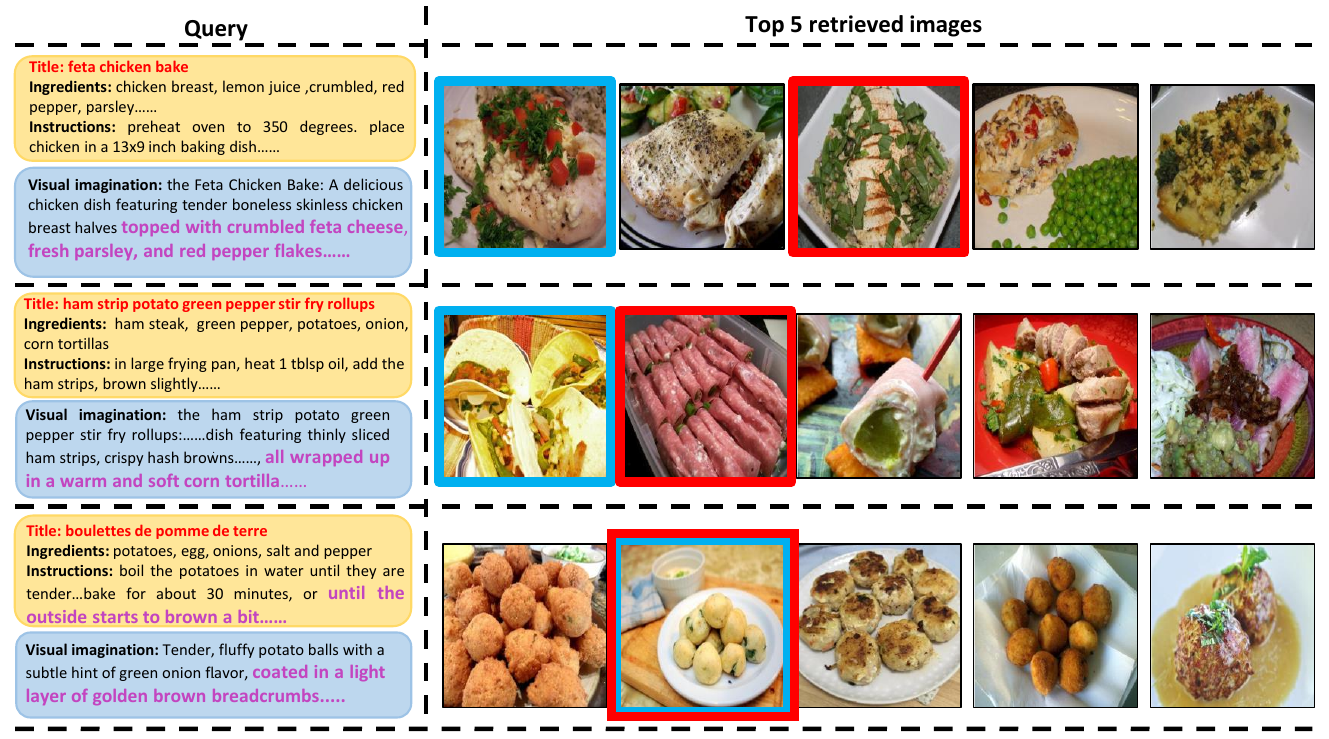}
    % \captionsetup{justification=centering}
    \caption{ Qualitative examples of recipe-to-image retrieval. The query in the left column shows the recipe and the corresponding visual imagination description produced by LLM. The right column shows the retrieved Top-5 food images from DAR++, where blue boxes represent the ground truth. We also highlight the Top-1 retrieved results of DAR with red bounding boxes.}
    %\vspace{-0.7cm}
    \label{fig:visual_r2i}
\end{figure}

% Use figure* for multi-column figure
\begin{figure}[t]
    \centering
    \includegraphics[width=0.86\linewidth]{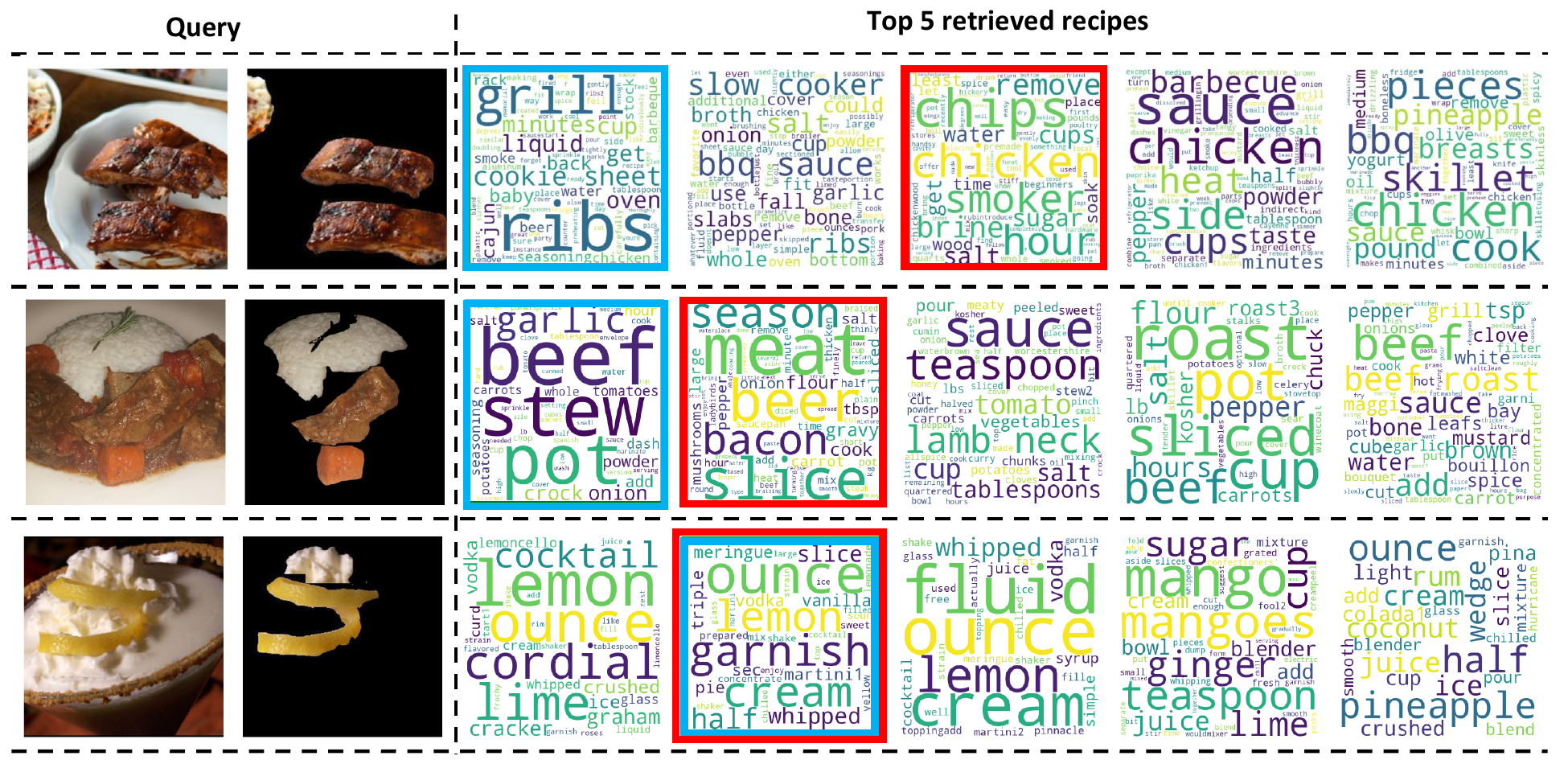}
    % \captionsetup{justification=centering}
    \caption{ Qualitative examples of image-to-recipe retrieval. The first two columns are image query and segments from SAM. The DAR++'s Top-5  retrieved recipes are represented by word clouds. The blue
boxes represent the ground truth and red boxes represent the Top-1
retrieved recipe of DAR.}
%\vspace{-0.5cm}
    \label{fig:visual_i2r}
\end{figure}

%\vspace{-0.35cm}
\subsection{Qualitative Analysis}
\label{subsection:Qualitative Analysis}
Fig.~\ref{fig:visual_r2i} and Fig.~\ref{fig:visual_i2r} show the qualitative examples of recipe-to-image retrieval and image-to-recipe retrieval respectively. The results are reported based on DAR++, which includes extra augmented image segments and visual imagination description for testing.

% \vspace{-20pt}
% \vspace{-0.2cm}
First, we show the result of retrieving the corresponding image using recipe as query in Fig.~\ref{fig:visual_r2i}. During inference, we can generate visual imagination description for each recipe query, which is combined as query to retrieve food images. In the first example, the query recipe is ``feta chicken bake''. We can see that all the Top-5 retrieval images are chicken, which are semantically and visually similar. Though our model DAR ranks the corresponding image in third position, our model DAR++ manages to improve the rank to first place. The reason is that our visual imagination description is capable of providing auxiliary visual-related information ``topped with fresh parsley'' and ``red pepper flakes ...'' to further distinguish the similar image returned by DAR but where the chicken is topped with scallions. The same phenomenon can be observed in the second example as well where ``rollups'' is interpreted as ``...wrapped up in corn tortilla'' to correct the result that retrieves a picture of the rollups made of meat by DAR. As the visual imagination description contains visual cues that do not exist in recipes, the recipe-to-image retrieval could be more interpretable by incorporating visual imagination description in the retrieval system. In the third example ``boulettes de pomme de terre'',  which means ``potato dumplings'', compared to DAR, our DAR++ is not able to rank higher than DAR, it is because the visual imagination provides misleading information ``coated in a light layer of golden brown breadcrumbs'' whereas in the recipe the color is described as ``brown a bit'' and there are no ingredients for ``breadcrumbs''.

As shown in Fig.~\ref{fig:visual_i2r}, we present examples of Top-5 retrieved recipes using food image as query of DAR++. Following~\cite{salvador2021revamping}, the recipes are shown with word clouds for better visualization. Given a query image, our DAR++ employs SAM to obtain image segments, which are subsequently combined as query as well. Multiple segments are combined as one image for visualization purpose. It can be observed that there is a correspondence between the segments and key ingredients in the corresponding recipe. For instance, as the most notable ingredients, ``ribs'' in the first example and ``beef'', ``carrots'' in the second example can be found in the segment picture and the first ranked retrieved recipes. As a result, our method potentially provides better interpretability for cross-modal recipe retrieval with the image segments.
% \noindent \textbf{Discussion about segments.} 
% Indeed, the quality of ingredient segments for augmentation relies heavily on the segmentation results generated by SAM. 

\noindent\textbf{Limitation of image segments.} Though image segments produced by SAM manage to boost the retrieval performance in both training and testing time, 
we are aware of the limitations as well. 
Similar to~\cite{zhang2023comprehensive,ji2023segment},
we also note that 
% due to the limitations of SAM ~\cite{zhang2023comprehensive,ji2023segment}, there is a possibility that 
SAM may miss small and irregular objects. Moreover, the ``everything'' mode samples a large number of point prompts for automatic segmentation, which results in SAM sometimes over-segmenting the fine-grained features of large and complex objects while neglecting their wholeness. In the third example in Fig.~\ref{fig:visual_i2r}, the ``cream'' with a large area is missing from the segments, which causes DAR++ to focus on the ``lemon'', thus DAR++ fails to rank the ground-truth recipe in the first place. Please see supplementary material for more details.
% In Fig.~\ref{fig:segment_hand}, SAM fails to segment the ``soda'' object against the complex background and ignores the overall ``cake'' by over-segmenting to the decorations on top. And after we manually improved the segments of these samples (in the third column), DAR++ with noiseless segments retrieved the correct recipe. 
% In summary, the introduction of segments provides fine-grained alignment information for retrieval, as well as more more interpretability for food images. However, at this stage, SAM still suffers from the above problems prevalent in segmentation models, which limits its augmentation effect.

%\vspace{-0.3cm}
\section{Conclusion}
%\vspace{-0.3cm}
% \textcolor{red}{
We have presented a new data augmentation paradigm for cross-modal recipe retrieval via foundation models. Leveraging the augmented data from SAM and Llama2, we propose DAR framework, which achieves state-of-the-art performance on Recipe1M dataset. We introduce lightweight adapter layers in CLIP to encode the original and augmented recipe and image data. Furthermore, we propose multi-level circle loss to perform multi-objective optimization to regularize the common embedding space. 
% Through ablation study, we show the effectiveness of adapter layers injecting into CLIP for cross-modal recipe retrieval with much less trainable parameters. The circle loss is shown to be more effective than triplet loss.
Importantly, we demonstrate that augmented data can not only be beneficial during training, but also can be used for test-time augmentation. While encouraging, it is inevitable to generate imperfect image segments which could limit the boost of cross-modal retrieval performance,
% noisy data from foundation models due to the hallucination problem of LLM and tendency of SAM to ignore small structures, 
which will be our future work.
% }

\textbf{Acknowledgement.} This work was partially supported by the National Natural Science Foundation of China (No. 62101524 and No. 62202439), and by the Singapore Ministry of Education (MOE) Academic Research Fund (AcRF) Tier 1 grant (No. MSS23C018). It was also supported by the advanced computing resources provided by the Supercomputing Center of the USTC.

\par\vfill\par

\bibliographystyle{splncs04}
\bibliography{main}
\setcounter{section}{0}
\renewcommand{\thesection}{\Alph{section}}

\section{Circle Loss}
We rewrite Eq.\ref{eq:circle_loss} to illustrate the flexibility and applicability of circle loss as follows:
\begin{equation}
L_{\text{circle}}(A,B) = \log [ 1 + {\textstyle \sum_{j=1}^{L}} e^{\gamma [ c_{n}^j + m ]_+ \cdot ( c_{n}^j - m ) } {\textstyle \sum_{i=1}^{K}} e^{\gamma [ 1+m- c_{p}^i]_+ \cdot(1-m-c_{p}^i)} ],
\label{eq:circle_loss}
\tag{6}
\end{equation}
where $m \in [0,1]$ represents the relaxation factor, and $\gamma$ is the scale factor used to rescale the similarity.
Circle loss aims to optimize $c_n \to 0$ and $c_p \to 1$ to push the distance between negative and positive pairs. With the relaxation factor $m$ to control the radius of the decision boundary, circle loss expects $c_n^j<m$ and $c_p^i>1-m$. In addition, in Eq.\ref{eq:circle_loss}, $\gamma [ c_{n}^j + m ]_+$ and $\gamma [ 1+m- c_{p}^i ]_+$ are multiplied as the weight factors on the similarity score of the negative and positive pairs. Taking $\gamma [ 1+m- c_{p}^i]_+ \cdot(1-m-c_{p}^i)$ as an example, if $c_p^i$ plus $m$ is still much less than the target optimization value of 1, $c_p^i$ will get a large weight for gradient updating during optimization, and the same for $c_n^j$. This allows each positive and negative pair to be optimized flexibly. 
% In the face of numerous embeddings, both original and augmented, circle loss is undoubtedly more capable of constructing complex common semantic spaces compared to triplet loss. 

\section{Additional Experiments}
\label{sec:add_exp}

\subsection{Trainable Parameters}
The adoption of adapter's structure allowed us to leverage CLIP's cross-modal retrieval capabilities with few parameters. In Table~\ref{tab:sup_params}, we provide a comparison of the number of trainable parameters with existing CLIP-based methods~\cite{shukor2022transformer, shukor2023vision}. In contrast to the full fine-tuning of CLIP, our DAR manages to achieve superior performance with much fewer parameters, showcasing the efficiency of our approach.
\subsection{Loss Weight Factors}
In the paper, we propose multi-level circle loss~\cite{sun2020circle} to regularize the common semantic space. Meanwhile, in order to balance the weights between each loss, we set the weight factors $\alpha$, $\beta$ and $\sigma$ for $L(E_S,E_R)$, $L(E_V,E_D)$ and $L_{\text{rec}}$ respectively. $\sigma$ is set to 1 following~\cite{salvador2021revamping}. Table~\ref{tab:sup_loss_weight} shows the performance of the model with different settings of $\alpha$ and $\beta$. The performance of assigning $\alpha=1, \beta=1$ is slightly better than that of $\alpha=0.1, \beta=0.1$, in particular in terms of R@1. In addition, we show more selection results on the hyperparameters $\gamma, m$ of circle loss in Table~\ref{tab:sup_loss_hyperparam}, which are searched on the validation set.
\subsection{Test-time Augmentation} 
% The augmented data in our method not only in the training phase but also in the test phase thus obtaining better performance, which can be used in other methods as well. Here 
We further examine the performance of H-T(ViT)~\cite{salvador2021revamping} with our proposed test-time augmentation by directly introducing the distance matrices of augmented data in DAR+ and DAR++ into its evaluation. As shown in the Table~\ref{tab:sup_eval}, similar to DAR+ and DAR++,  H-T(ViT)+ and  H-T(ViT)++ exhibit consistent improvements over H-T(ViT), which further validate the effectiveness of test-time augmentation.
% \usepackage{booktabs}
% \usepackage{multirow}
\begin{table}[]
\centering

\caption{ Comparison with the number of trainable parameters for the method that also uses CLIP. Size is in millions, and ``Our Proportion'' represents the percentage of our parameters size compared to the corresponding methods.}
\begin{tabular}{@{}ccc@{}}
\toprule
\multirow{2}{*}{\textbf{Methods}} & \multicolumn{2}{c}{\textbf{Trainable Params}}   \\ \cmidrule(l){2-3} 
                                  & \textbf{Size (M)} & \textbf{Our Proportion (\%)} \\ \midrule
T-Food (CLIP-ViT)~\cite{shukor2022transformer}                  & 392.6            & 11.1                         \\
VLPCook~\cite{shukor2023vision}                           & 314.6            & 13.8                         \\
\textbf{DAR}                               & 43.6             & 100                          \\ \bottomrule
\end{tabular}

\label{tab:sup_params}

\end{table}
% \usepackage{booktabs}
% \usepackage{multirow}
\begin{table}[]
\centering
\caption{ Performance comparison of multi-circle loss with different loss weight of $(\alpha,\beta)$.}
\begin{tabular}{@{}ccccccccc@{}}
\toprule
\multirow{2}{*}{\textbf{Weight Factor Setting}} & \multicolumn{4}{c}{\textbf{image-to-recipe}}                & \multicolumn{4}{c}{\textbf{recipe-to-image}}                \\ \cmidrule(l){2-9} 
                                                     & \textbf{medR} & \textbf{R@1} & \textbf{R@5} & \textbf{R@10} & \textbf{medR} & \textbf{R@1} & \textbf{R@5} & \textbf{R@10} \\ \midrule
$\alpha=0.1,\beta=0.1$                                              &      2.0         &  43.3            &    72.7         &  82.2             &    2.0           &     44.0         &       73.5       &       82.9        \\

$\alpha=1 , \beta=1$                                                &\textbf{2.0}           & \textbf{44.2}         & \textbf{73.2}         & \textbf{82.4}          & \textbf{2.0}           & \textbf{44.8}         & \textbf{73.9}         & \textbf{83.1}          \\ \bottomrule
\end{tabular}

\label{tab:sup_loss_weight}

\end{table}
% \usepackage{booktabs}
% \usepackage{multirow}
\begin{table}[]
\centering
\caption{ Performance comparison of circle loss  with different hyperparameter $\gamma, m$ under baseline setting.}
\begin{tabular}{@{}ccccccccc@{}}
\toprule
\multirow{2}{*}{\textbf{Hyperparameters Setting}}                                 & \multicolumn{4}{c}{\textbf{image-to-recipe}}                & \multicolumn{4}{c}{\textbf{recipe-to-image}}                \\ \cmidrule(l){2-9} 
                                                                 & \textbf{medR} & \textbf{R@1} & \textbf{R@5} & \textbf{R@10} & \textbf{medR} & \textbf{R@1} & \textbf{R@5} & \textbf{R@10} \\ \midrule
$\gamma=32, m=0.2$                                    & 2.0           & 41.6         & 71.1         & 81.2          & 2.0           & 42.3         & 71.8         & 81.7          \\
 $\gamma=24, m=0.25$                                   & 2.0           & 41.1         & 70.6         & 80.5          & 2.0           & 42.4         & 71.4         & 81.1          \\
 $\gamma=32, m=0.25$ \textbf{(paper)} & \textbf{2.0}  & \textbf{42.3} & \textbf{71.8} & \textbf{81.5} & \textbf{2.0}  & \textbf{43.4} & \textbf{72.8} & \textbf{82.3} \\ \bottomrule
\end{tabular}

\label{tab:sup_loss_hyperparam}

\end{table}
% \usepackage{booktabs}
% \usepackage{multirow}
\begin{table}[]
\centering
\caption{ Results of adding data augmentation to  H-T(ViT)~\cite{salvador2021revamping} model at test-time.}
\begin{tabular}{@{}ccccccccc@{}}
\toprule
\multirow{2}{*}{\textbf{Evaluation Method}} & \multicolumn{4}{c}{\textbf{image-to-recipe}}                  & \multicolumn{4}{c}{\textbf{recipe-to-image}}                  \\ \cmidrule(l){2-9} 
                                            & \textbf{medR} & \textbf{R@1}  & \textbf{R@5}  & \textbf{R@10} & \textbf{medR} & \textbf{R@1}  & \textbf{R@5}  & \textbf{R@10} \\ \midrule
H-T(ViT)                                    & 3.0           & 33.5          & 62.1          & 72.8          & 3.0           & 33.7          & 62.2          & 72.7          \\
H-T(ViT)+                                   & 2.0           & 41.2          & 70.5          & 80.2          & 2.0           & 41.2          & 69.7          & 79.5          \\
H-T(ViT)++                                  & \textbf{2.0}  & \textbf{42.4} & \textbf{71.8} & \textbf{81.4} & \textbf{2.0}  & \textbf{41.9} & \textbf{70.7} & \textbf{80.3} \\ \bottomrule
\end{tabular}

\label{tab:sup_eval}

\end{table}
\section{Visualization and Analysis }
In this section, we provide more qualitative examples and result analysis.
% we show more examples of data augmentation to highlight its effectiveness in modal alignment. At the same time, we also find that the foundation model brings a certain amount of noise samples into the data due to its own limitations. We analyze the problem and propose ideas for further improvements in the future.
\subsection{Visual Imagination Description by LLM}
% or we design new prompt engineering to allow the LLM  to determine whether there are illusions in the description.
% Use figure* for multi-column figure
\begin{figure}[]
    \centering
    \includegraphics[width=\textwidth ]{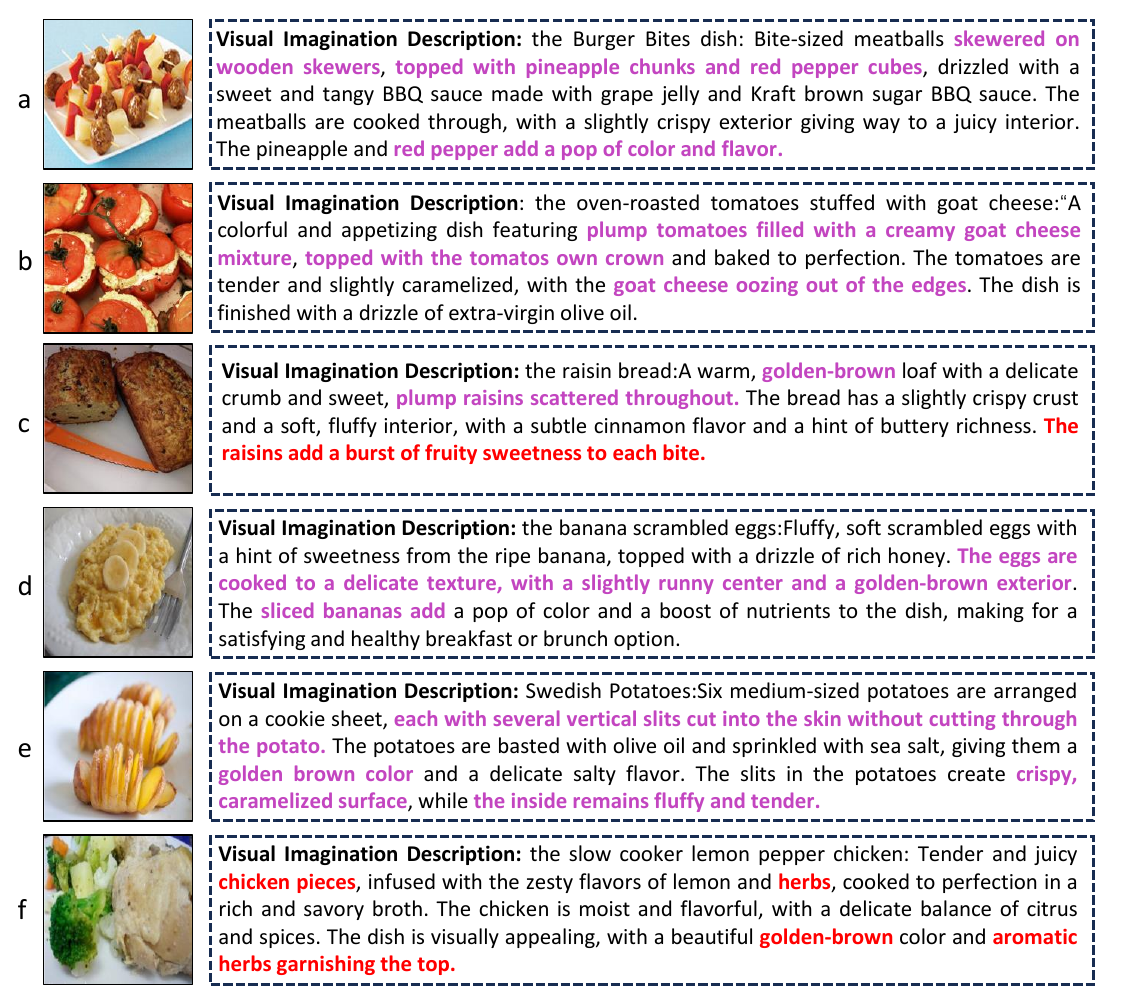}
    % \captionsetup{justification=centering}
    \caption{Examples of visual imagination descriptions and corresponding images. Note that the visual imagination descriptions are generated based on recipes.}
    \label{fig:sup_visual_r2i}
\end{figure}

% Use figure* for multi-column figure
\begin{figure}[]
    \centering
    \includegraphics[width=\textwidth]{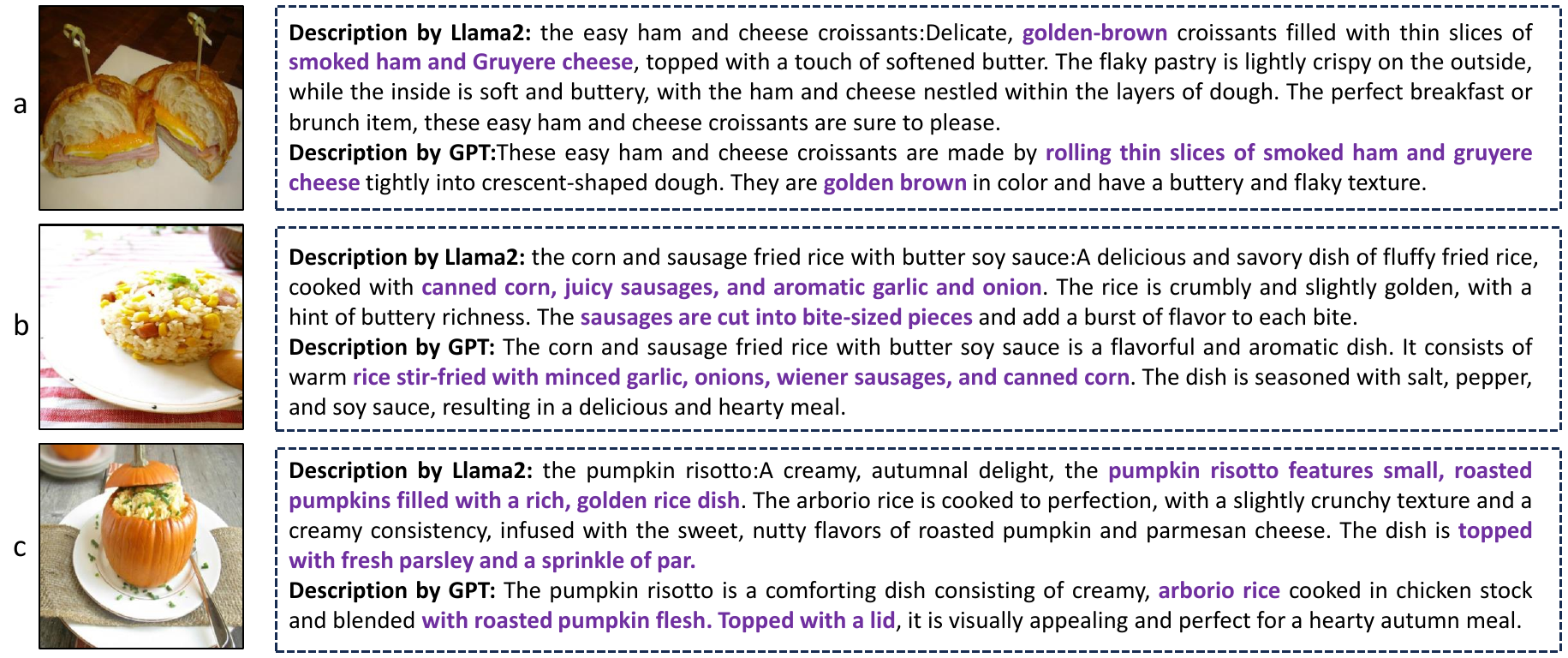}
    % \captionsetup{justification=centering}
    \caption{ Comparison of examples of visual imagination descriptions generated by different LLMs. The figure includes visual imagination descriptions generated by Llama2 and GPT-3.5 about the same recipes and the corresponding images.}
    \label{fig:sup_visual_gptcompare}
\end{figure}

In Fig~\ref{fig:sup_visual_r2i}, we show more samples of visual imagination descriptions and corresponding image, notably ``goat cheese oozing out of the edges'' of example (b) and ``each with several vertical slits cutting into the skin without cutting through the potato'' of example (e) and other fine-grained alignment information.

In addition to Llama2 adopted in the main paper, we also examine GPT-3.5 to generate visual imagination description. We show the visual imagination description obtained by Llama2 and GPT-3.5 for the same recipe with the same prompt in Fig.~\ref{fig:sup_visual_gptcompare}.
Here we adopt  \textit{gpt-3.5-turbo} model for GPT-3.5 and  \textit{Llama2-13b-chat} model for
Llama2.  
In the example in Fig.~\ref{fig:sup_visual_gptcompare}(a), both models capture the ``golden-brown'' color and the ``filled with smoked ham and Gruyere cheese'' feature. In (b), our Llama2 also focuses on the detail that ``sausages are cut into bite-sized pieces''. In contrast to the GPT-3.5 which only notes ``Topped with a lid'' of the pumpkin risotto in (c), our model describes it as ``pumpkins filled with a rich, golden rice''. As a result, both LLMs capture key visual features in the recipe. 
%These results indicate that our results generated by the Llama2  are competitive. 

% However, in example (c), the description of ``The raisins add a burst of fruity sweetness to each bite'' is a flavor feature that is not presented in the original recipe. In addition, the information of ``herbs'', ``chicken pieces'' and ``aromatic herbs garnishing the top''  are all LLM hallucinations~\cite{ji2023survey,rawte2023survey}. We infer that one of the reasons for the hallucination is that LLM has over-inference on the recipe information, and some irrelevant or even misleading features are being amplified by the model.
% Another reason is that although we restrict LLM from utilizing external information, it inevitably combines information from its own knowledge base, which may have some deviation from the original recipe. 
% In future work, we plan to filter noise samples by comparing the similarity between the recipes and descriptions and regenerate them, alternatively, we can use different prompts to generate several visual imagination descriptions to make them more robust.

% Use figure* for multi-column figure
\begin{figure}[]
    \centering

    \includegraphics[width=0.85\textwidth]{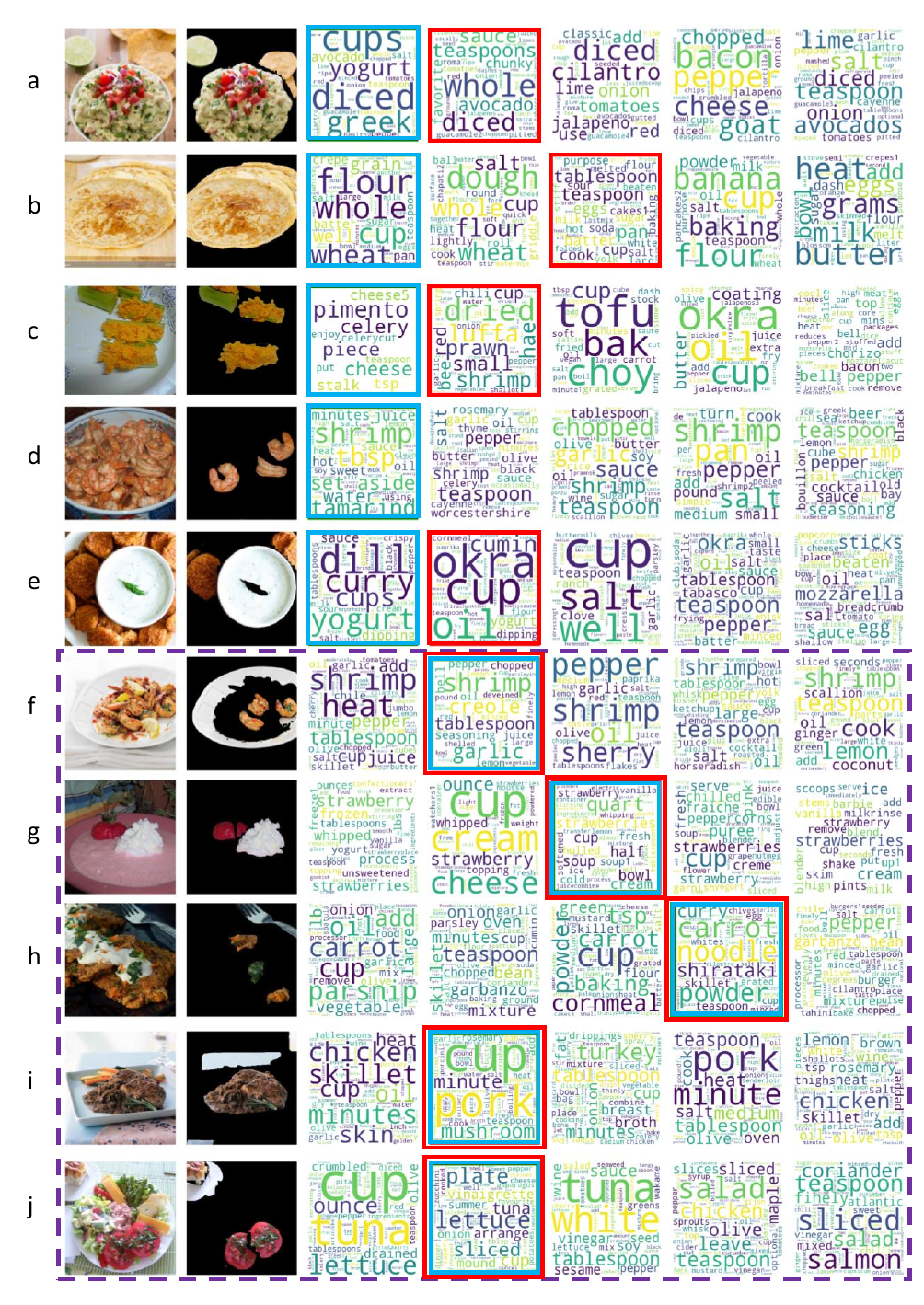}
    % \captionsetup{justification=centering}
    \caption{ Qualitative examples of image-to-recipe retrieval. The first two columns are image query and segments from SAM. The DAR++'s Top-5  retrieved recipes are represented by word clouds. The blue
boxes represent the ground truth and red boxes represent the Top-1
retrieved image of DAR. The large purple box includes samples of poor quality segments.}
    \label{fig:sup_visual_i2r}
\end{figure}

% Use figure* for multi-column figure
\begin{figure}[]
    \centering

    \includegraphics[width=0.9\textwidth]{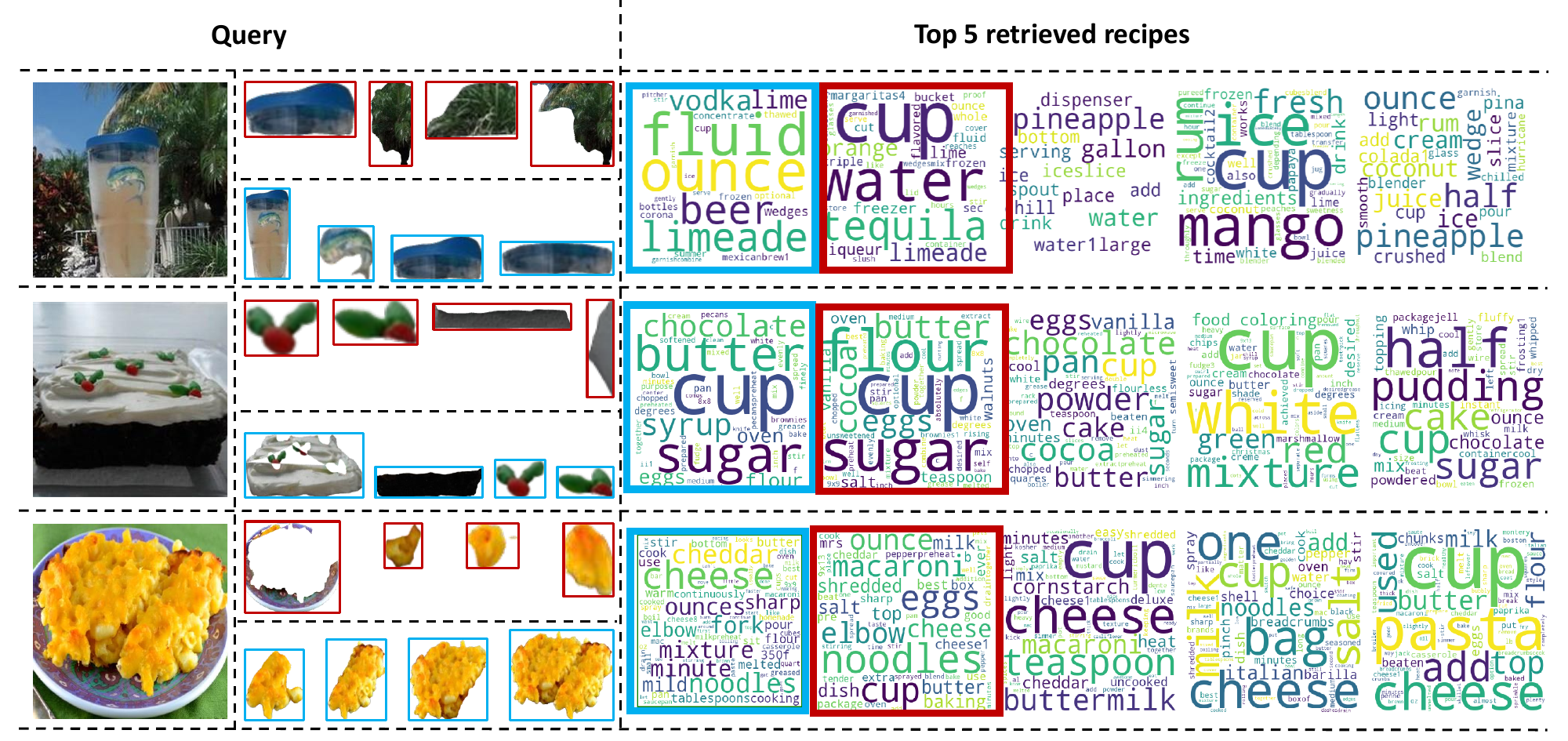}
    % \captionsetup{justification=centering}
    \caption{ Qualitative examples of the effect of SAM imperfections on retrieval. The first two columns are the image query and segments from SAM, where the column of segments is divided into two rows. The upper segments with red boxes are automatically generated by SAM while the lower segments with blue boxes are hand-picked using SAM.  The DAR++'s Top-5  retrieved recipes with hand-picked segments are represented by word clouds. The blue
boxes represent the ground truth. The red boxes represent the Top-1 retrieved image of DAR++ with automatically generated segments.}
    \label{fig:sup_seg_change}
    
\end{figure}

\subsection{Image Segments by SAM}
Fig.~\ref{fig:sup_visual_i2r} shows more qualitative examples of image segments and the corresponding word clouds of top-5 retrieved recipes. 
% To correspond to the original image, segments are drawn on a single image. 
It can be seen that SAM accurately segments the food parts from the original image, e.g., in example (b), the pasty is accurately distinguished from the similarly colored table, and in example (d), the three shrimps clearly segmented correspond to the ``shrimp'' ingredient in the recipe. These results show that segments effectively help DAR++ to align the image and the recipe through the ingredients.

Nevertheless, previous experimental results in the main paper both in terms of the degradation of model performance after adding adapter layers into the segment encoder, and the fact that DAR++ cannot bring stable improvement compared to the DAR+ during test-time augmentation, we believe that these problems are due to the quality of some segments obtained by SAM. 
% There has been some works~\cite{ji2023segment,zhang2023comprehensive} analyzing the shortcomings of SAM. 
The inherent shortcomings of SAM make some segments mixed with noise~\cite{ji2023segment,zhang2023comprehensive}. We show some representative noise samples (examples with purple box) selected from the recipe1M in Fig.~\ref{fig:sup_visual_i2r} to illustrate the problems we observed in SAM segmentation for data augmentation. There are two imperfections of SAM, the first one is that it sometimes misses small, irregular structures, such as the omission of lemon in example (f) and the discarding of irregular parts of the salad in example (j). Another imperfection is that the ``everything'' mode over-segments the large objects that include many details and ignores overall semantics at times. In example (g), SAM ignores the whole cream at the bottom, while in (h) it ignores the whole pork chop. 
In addition, the filtering of segments is not successful for all samples, which results in background information remaining in examples (f) and (i). 

To address the problems of segments, we designed sample analysis experiments to illustrate the future implications of solving these problems. Specifically,  for a portion of the test samples that were retrieved incorrectly by DAR++, we would like to eliminate the noise in the segments by manually utilizing SAM (with points prompt) to obtain image segments instead of the original segments. The experimental results also show that the bad segment samples limit the further improvement of model performance. In Fig.~\ref{fig:sup_seg_change}, for the original segmentation, the ``Sprite'' in the first row is plagued by the complex background and only the lid of the cup is segmented. Due to SAM's over-segmentation of large objects, the cake in the second row and the cheese in the third row only retain the decoration on top of the cake and a very small portion of the cheese, and also contain irrelevant information such as the background and the plate. With our manual segments improving these problems, DAR++ successfully retrieved the correct recipes with blue boxes. We will explore how to further improve the quality of the segments in the future work.

% We will explore combine food-specific segmentation models~FoodSAM~\cite{} and 
% As a result, addressing these issues will make a lot of sense and become our future work. we will reduce the impact of small portions of noisy data by fine-tuning  SAM on the food images, or by building a more robust segment encoder in which different segments are weighted instead of averaged.

\end{document}